\documentclass[twocolumn,aps,prd,superscriptaddress,nofootinbib]{revtex4-1}
\usepackage{amssymb}
\usepackage{mathbbol}
\usepackage{amsfonts}
\usepackage{mathrsfs}
\usepackage{epsfig,bm,dcolumn}
\usepackage{graphicx}
\usepackage{color}
\usepackage{amsmath}
\usepackage{dcolumn}
\usepackage{overpic}
\usepackage[colorlinks, linkcolor=red,anchorcolor=green,citecolor=blue]{hyperref}
\usepackage{slashed}

\usepackage[normalem]{ulem}

\usepackage{indentfirst}
\usepackage{bm}

\usepackage{simplewick}

\newcommand{\pt}{\partial}

\newcommand{\Tr}{\mathrm{Tr}}

\graphicspath{{picture/}}
\usepackage{color}

\begin{document}
	\title{Dynamical magnetic fields in heavy-ion collisions}
	\author{Anping Huang}
	\affiliation{School of Nuclear Science and Technology, University of Chinese Academy of Sciences, Beijing 100049, China}
	\affiliation{Physics Department and Center for Exploration of Energy and Matter,	Indiana University, 2401 N Milo B. Sampson Lane, Bloomington, Indiana 47408, USA}
	\author{Duan She}
	\affiliation{Department of Modern Physics, University of Science and Technology of China, Anhui 230026, China}
	\affiliation{Key Laboratory of Quark and Lepton Physics (MOE), Central China Normal University, Wuhan 430079, China}
	\author{Shuzhe Shi} \email{shuzhe.shi@stonybrook.edu}
	\affiliation{Department of Physics and Astronomy, Stony Brook University, Stony Brook, New York 11794-3800, USA}
	\author{Mei Huang} \email{huangmei@ucas.ac.cn}
	\affiliation{School of Nuclear Science and Technology, University of Chinese Academy of Sciences, Beijing 100049, China}
	\author{Jinfeng Liao}\email{liaoji@indiana.edu}
	\affiliation{Physics Department and Center for Exploration of Energy and Matter,	Indiana University, 2401 N Milo B. Sampson Lane, Bloomington, Indiana 47408, USA}
	\date{\today}

\begin{abstract}
The magnetic fields in heavy-ion collisions are important ingredients for many interesting phenomena, such as the Chiral Magnetic Effect, Chiral Magnetic Wave, the directed flow $v_1$ of $D^0$ mesons and the splitting of the spin polarization of the $\Lambda$/$\bar{\Lambda}$. Quantitative studies of these phenomena however suffer from limited understanding on  the dynamical evolution of these fields in the medium created by the collisions, which remains a critical and challenging problem. The initial magnetic fields from the colliding nuclei   decay very fast in the vacuum but their  lifetime could be extended through medium response due to electrically conducting quarks and antiquarks. Here we perform a detailed analysis of  such medium effect on the dynamical magnetic fields  by numerically solving the Maxwell's equations  concurrently  with the  expanding medium described by viscous hydrodynamics, under the assumption of  negligible back reaction of the fields on the fluid evolution. Our results suggest a considerable enhancement of late time magnetic fields, the magnitude of which depends sensitively on the fireball expansion as well as the medium electric conductivity both before and during hydrodynamic stage.
\end{abstract}
	
\maketitle

\section{Introduction}

The ultra-relativistic heavy-ion collisions provide the opportunity to create quark-gluon plasma (QGP) and  investigate its properties  under extreme conditions in terms of temperatures, baryon densities, and more recently  also  magnetic fields and vorticity. There are very strong magnetic fields arising from the fast-moving ions in the non-central heavy-ion collisions, which can reach about $eB\sim\,m^{2}_{\pi}\sim\,10^{18}$~Gauss in Au+Au collisions at the Relativistic Heavy Ion Collider (RHIC), and can be still an order of magnitude larger at the Larger Hadron Collider (LHC)~\cite{Kharzeev:2007jp,Skokov:2009qp,Voronyuk:2011jd,Bzdak:2011yy,Deng:2012pc,Bloczynski:2012en,McLerran:2013hla,Tuchin:2015oka,Chen:2021nxs}.
Many interesting effects induced by such magnetic fields have been proposed and studied both theoretically and experimentally, such as the Chiral Magnetic Effect, Chiral Magnetic Wave, the directed flow $v_1$ of $D^0$ mesons and the splitting of the spin polarization of the $\Lambda$/$\bar{\Lambda}$, etc. See recent reviews in e.g.~\cite{Kharzeev:2012ph,Kharzeev:2015znc,Kharzeev:2020jxw,Fukushima:2018grm,Shovkovy:2021yyw,Li:2020dwr,Huang:2015oca}. 

While the initial   strength and spatial distribution of the magnetic fields at the beginning of a heavy ion collision can be accurately calculated,  the subsequent dynamical evolution of such magnetic fields in the medium is rather poorly determined. If one only considers the field evolution in vacuum case,  it is well known that the strength decays rapidly in time   and the field lifetime at mid-rapidity can be estimated as $\tau_{B}\sim\,R_{A}/(\gamma\,v_{z})$ which is about $0.06$~fm for Au+Au collision at $200$~GeV~\cite{Huang:2015oca} while  about $0.005$~fm for Pb+Pb collions at $2.76$~TeV. However, the lifetime of the in-medium magnetic field could be elongated due to the presence of the quark-gluon plasma  in which the electrically charged quarks and anti-quarks form a conducting medium with induction effect, as qualitatively demonstrated by various theoretical and numerical investigations~\cite{Gupta:2003zh,Aarts:2007wj,Ding:2010ga,Ding:2016hua,Aarts:2002tn,Huang:2013iia,Jiang:2014ura,Bannur:2006js,Das:2019wjg,Hosoya:1983xm,Nam:2012sg,Cassing:2013iz,Arnold:2003zc,Wang:2021oqq}.  
A quantitative understanding of  the dynamical evolution of magnetic fields, however, remains a key challenge.

Generally speaking, there are two different types of approaches, the ``strong field'' and ``weak field'' methods. In the strong field method, the  influence of the electromagnetic fields on the medium evolution can not be ignored and thus need to be taken into account for describing the medium. The most representative example is the Magneto-hydrodynamics (MHD). In an ideal MHD with infinite conductivity, the magnetic field obeys the frozen flux (or Alfven) theorem and can therefore be represented simply in time~\cite{Roy:2015kma, Pu:2016ayh} for a Bjorken flow, i.e $B(\tau)=B_0\, \tau_{0}/\tau$, where $B_{0}$ is the initial magnetic field at time $\tau_{0}$. Numerical efforts were developed in~\cite{Inghirami:2016iru, Inghirami:2019mkc} by using the improved version of ECHO-QGP to simulate the evolution of electromagnetic fields in the heavy-ion collisions by solving the relativistic ideal MHD equations   with the assumption of infinite electrical conductivity of the plasma and the ideal hydro for the medium without the dissipative effects. These analyses show   that the medium effect would indeed slow down the decay of the magnetic field and hence enlarge its lifetime. However, given the strongly coupled nature of the quark-gluon plasma, it is difficult to imagine that the electric conductivity would be very large. In fact, lattice simulations would suggest a rather limited QGP electric conductivity. 

In the weak field method, one assumes that  the effect of the medium on electromagnetic fields must be accounted for while the back reaction of electromagnetic fields on the medium is negligible.  In this approach, the medium evolution can be described by usual viscous hydrodynamics without electromagnetic fields and the evolution of the electromagnetic fields can be derived from the Maxwell's equations by including the responses from the medium via e.g. induction currents. 
Several previous theoretical and numerical works adopted this method, see e.g.~\cite{Tuchin:2014iua, Stewart:2017zsu, Stewart:2021mjz, Gursoy:2014aka,Gursoy:2018yai,Li:2016tel,Tuchin:2015oka,McLerran:2013hla,Zakharov:2014dia,Amato:2013oja}.   
These studies also clearly demonstrated the medium response effect that can help extend the lifetime of magnetic fields, but often suffer from various unrealistic approximations e.g. constant conductivity, static medium, 1D Bjorken expansion only, infinite transverse medium, etc.

Quantitatively understanding the dynamical evolution of magnetic fields requires a more realistic hydrodynamic  background, a more realistic QGP conductivity,  a proper treatment of the full spacetime dependence of the fields, as well as a careful analysis of the per-hydro non-equilibrium stage. In this work, we make an attempt at address these issues based on the weak field method through numerically solving concurrently the viscous hydrodynamics for the medium and the Maxwell's equations for the electromagnetic fields.  To be specific, let's   take the $\sqrt{s_\mathrm{NN}}=200$~GeV Au+Au collisions as an example to demonstrate the developed framework. The full evolution includes three different stages in our framework.  
The first stage is the initial stage of time interval $\tau=0.0 \sim~0.1$~fm, which may be gluon dominated with few quarks. At this time there would be no medium response and the electromagnetic fields are assumed to evolve in vacuum. The second stage is the pre-equilibrium stage of time interval $\tau=0.1 \sim 0.4$~fm where the system is undergoing Bjorken expansion. While substantial number of quarks and antiquarks emerge in this stage, they may or may not be close to thermal equilibrium yet. Since there is no clear answer for the electric conductivity in such non-equilibrium case, we will test several plausible assumptions for the pre-equilibrium effective electric conductivity, such as the zero model, constant model and linear model. 
The third stage is the hydrodynamic stage for time $\tau\geq0.4~\text{fm}$, in which the QGP medium is assumed to have thermal conductivity and several scenarios for the conductivity will also be tested. 
A more detailed description of our framework will be presented later. 

The rest of this paper is organized as follows. In Sec.~\ref{sec-1}, the analytical solution and the numerical algorithm of Maxwell's equations in  the static QGP scenario are briefly reviewed, with certain new results  for dynamic magnetic fields under static QGP with space-time-dependent conductivity. In Sec.~\ref{sec-2}, the dynamical evolution of magnetic fields in dynamically expanding QGP is studied and the results are compared for different models of both thermal and pre-equilibrium conductivities as well as for different choices of hydrodynamic backgrounds. Finally we conclude in Sec.~\ref{sec-3}.  A number of relevant technical details are also included in  several appendices: 
Appendix~\ref{app-1} presents the external electric and magnetic fields in heavy-ion collision as solutions of Eq.~\eqref{eq:meex};  Appendix~\ref{app-2} introduces the Yee-grid algorithm for solving the Maxwell's equations; Appendix~\ref{app-3} briefly reviews the Levi--Civita tensor and electromagnetic tensor in Milne space used for the expanding case;   Appendix~\ref{app-4} discusses   the difference of the velocity in Milne space and Minkowski space.

\section{Dynamical magnetic field in a Static QGP}\label{sec-1}
The covariant Maxwell equations are 
\begin{align}\begin{split}\label{eq:cme}
	&\partial_{\mu}F^{\mu\nu}=J^{\nu},\\
	&\partial_{\mu}\widetilde{F}^{\mu\nu}=0,
\end{split}\end{align}
where $\widetilde{F}^{\mu\nu}=\frac{1}{2}\epsilon^{\mu\nu\alpha\beta}F_{\alpha\beta}$ is a dual tensor of the electromagnetic field tensor $F^{\mu\nu}=\partial^{\mu}A^{\nu}-\partial^{\nu}A^{\mu}$. Using the relations $B^{i}=-\frac{1}{2}\epsilon^{ijk}F_{jk}=\widetilde{F}^{i0},\quad E^{i}=F^{i0},\quad F^{ij}=\epsilon^{ijk}B_{k}$, the covariant Maxwell equations~\eqref{eq:cme} can be rewritten as the familiar form,
\begin{align}\label{eq:me}
	\begin{split}
	&\triangledown\cdot\mathbf{E}=J^{0},\\
	&\triangledown\cdot\mathbf{B}=0,\\
	&\partial_{t}\mathbf{E}=\triangledown\times\mathbf{B}-\mathbf{J},\\
	&\partial_{t}\mathbf{B}=-\triangledown\times\mathbf{E}.
	\end{split}
\end{align}
The first two equations are the constraint equations, while the last two equations the dynamical equations of the electromagnetic fields. The latter can be used to derive the electric and magnetic fields at next time step. In a static medium, the current $J^{0}$ and $\mathbf{J}$ can expanded as the following,
\begin{align}
	J^{0}=J^{0}_{s},\qquad
	\mathbf{J}=\sigma\mathbf{E}+\sigma_{\chi}\mathbf{B}+\mathbf{J}_{s},
\end{align}
with $\sigma$ and $\sigma_{\chi}$ respectively being the electric and chiral conductivities of QGP. $J^{0}_{s}$ and $\mathbf{J}_{s}$ are the source contributions from the fast moving protons in the colliding nuclei. They can be written as
\begin{align}
	J^{0}_{s}=\;&
	e\sum_{i}\delta(\mathbf{x}_{\perp}-\mathbf{x}^{'}_{\perp,i})\delta(z-z^{'}_{i}-\beta t),\\
	\mathbf{J}_{s}=\;&
	e\sum_{i}\beta\hat{z}\,\delta(\mathbf{x}_{\perp}-\mathbf{x}^{'}_{\perp,i})\delta(z-z^{'}_{i}-\beta t).
\end{align} 

From the above Maxwell equations~\eqref{eq:me}, we can construct the corresponding wave equations for the electric and magnetic fields,  
\begin{align}\begin{split}\label{eq:we}
&   \left(\triangledown^{2}-\partial_{t}^{2}-\sigma\partial_{t}\right)\mathbf{E}-\sigma_{t}\mathbf{E}+\sigma_{\chi}\triangledown\times\mathbf{E}
\\=\;&
    \sigma_{\chi,t}\,\mathbf{B}+\partial_{t}\mathbf{J}_{s}+\triangledown J^{0}_{s},\\~\\
&    \left(\triangledown^{2}-\partial_{t}^{2}-\sigma\partial_{t}\right)\mathbf{B}+(\triangledown\sigma_{\chi})\times\mathbf{B}+\sigma_{\chi}\triangledown\times\mathbf{B}
\\=\;&
    -(\triangledown\sigma)\times\mathbf{E}-\triangledown\times\mathbf{J}_{s},
\end{split}\end{align}
where we have used the notations $\sigma_{t}=\partial_{t}\sigma$, $\sigma_{\chi,t}=\partial_{t}\sigma_{\chi}$.

\subsection{analytical solution of Maxwell Equations at constant conductivities}
It is possible to analytically solve the Maxwell equations Eq.~\eqref{eq:me} or Eq.~\eqref{eq:we}, when the electric and chiral conductivity are all constant for space-time~\cite{Tuchin:2014iua, Gursoy:2014aka,Gursoy:2018yai,Li:2016tel}. In such a case, the wave equation~\eqref{eq:we} can be simplified as
\begin{align}
\begin{split}
&\left(\triangledown^{2}-\partial_{t}^{2}-\sigma\partial_{t}\right)\mathbf{E}+\sigma_{\chi}\triangledown\times\mathbf{E}=\partial_{t}\mathbf{J}_{s}+\triangledown J^{0}_{s},\\
&\left(\triangledown^{2}-\partial_{t}^{2}-\sigma\partial_{t}\right)\mathbf{B}+\sigma_{\chi}\triangledown\times\mathbf{B}=-\triangledown\times\mathbf{J}_{s}.
\end{split}
\end{align}
Adopting the cylindric coordinate, its analytical solution is found to be~\cite{Tuchin:2014iua, Gursoy:2014aka,Gursoy:2018yai,Li:2016tel}
\begin{align}
\begin{split}
B_{\phi} =\;& 
    \frac{Q}{4\pi} \frac{v\gamma x_{T}}{\Delta^{3/2}}\left(1+\frac{\sigma v\gamma}{2}\sqrt{\Delta}\right)e^{A},  \\
B_{r} =\;&
    -\sigma_{\chi}\frac{Q}{8\pi} \frac{v\gamma^{2}x_{T}}{\Delta^{3/2}} \left[\gamma(vt-z)+A\sqrt{\Delta}\right]e^{A},  \\
B_{z} =\;&
    \sigma_{\chi}\frac{Q}{8\pi} \frac{v\gamma}{\Delta^{3/2}} \Big[\gamma^{2}(vt-z)^{2}\left(1+\frac{\sigma v\gamma}{2}\sqrt{\Delta}\right) 
\\&\quad
    + \Delta\left(1-\frac{\sigma v\gamma}{2}\sqrt{\Delta}\right)\Big] e^{A},
\end{split}\label{Eq:inB}
\\~\nonumber\\
\begin{split}
E_{\phi}  =\;& 
	\sigma_{\chi}\frac{Q}{8\pi}\frac{v^{2}\gamma^{2}x_{T}}{\Delta^{3/2}}\left[\gamma(vt-z)+A\sqrt{\Delta}\right]e^{A},\\
E_{r}  =\;& 
	\frac{Q}{4\pi}\bigg\{ \frac{\gamma x_{T}}{\Delta^{3/2}}\left(1+\frac{\sigma v\gamma}{2}\sqrt{\Delta}\right)
\\&\quad
	-\frac{\sigma}{vx_{T}}e^{-\sigma(t-z/v)}\left[1+\frac{\gamma(vt-z)}{\sqrt{\Delta}}\right]\bigg\} e^{A}, \\
E_{z}  =\;& 
	\frac{Q}{4\pi}\bigg\{ -e^{A}\frac{1}{\Delta^{3/2}}\left[\gamma(vt-z)+A\sqrt{\Delta}+\frac{\sigma\gamma}{v}\Delta\right]
\\&\quad
	+\frac{\sigma^{2}}{v^{2}}e^{-\sigma(t-z/v)}\Gamma(0,-A)\bigg\},
\end{split}\label{eq:e-phi}
\end{align}
where $\Delta\equiv\gamma^{2}(vt-z)^{2}+x_{T}^{2}$, $A\equiv(\sigma v\gamma/2)[\gamma(vt-z)-\sqrt{\Delta}]$, $\Gamma(0,-A)$ is the incomplete gamma function defined as
$\Gamma(a,z)=\int_{z}^{\infty}dt\, t^{a-1}e^{-t}$. 

As noted above, such an analytical solution is based on the precondition that the electric and chiral conductivities are space-time independent. Such a condition is not satisfied by the rapidly expanding medium form in the heavy-ion collision.
It seems unrealistic to analytically solve the wave equations Eq.~\eqref{eq:we} or the Maxwell equations Eq.~\eqref{eq:me} when the conductivities are space-time dependent. To further investigate the realistic dynamical evolution of the electromagnetic field in heavy-ion collisions, it calls for the numerical calculations.

\subsection{numerical method to solve Maxwell Equations}\label{sec:NME}
Numerically solving the Maxwell equations in Eq.~(\ref{eq:me}) might be unstable, due to the Dirac delta functions in the source term. Therefore, we will adopt the method established by McLerran and Skokov~\cite{McLerran:2013hla}. In this method, the electric and magnetic fields are separated into two pieces, i.e 
\begin{align}
	&\mathbf{E}=\mathbf{E}_{ext}+\mathbf{E}_{int},\qquad \mathbf{B}=\mathbf{B}_{ext}+\mathbf{B}_{int}.
\end{align} 
The subscript ``ext'' denotes the external part which originated by the source contribution from the fast moving charge particles in heavy-ion collisions, whereas ``int'' refers to the induced electromagnetic fields generated in the created quark-gluon plasma (QGP). Then the Maxwell equations in Eq.~\eqref{eq:me} under static medium now can been split into two parts. For the ``external'' part,
\begin{align}\label{eq:meex}
\begin{split}
&\triangledown\cdot\mathbf{E}_{ext}=J^{0}_{s},\\
&\partial_{t}\mathbf{E}_{ext}=\triangledown\times\mathbf{B}_{ext}-\mathbf{J}_{s},\\
&\triangledown\cdot\mathbf{B}_{ext}=0,\\
&\partial_{t}\mathbf{B}_{ext}=-\triangledown\times\mathbf{E}_{ext}.
\end{split}
\end{align}
There is analytical solution to this set of equations, which is the electric and magnetic fields induced by the fast moving charged particles. Details can be found in Appendix~\ref{ap:EB-moving-ion}. The ``internal'' part is,
\begin{align}\label{eq:meint}
\begin{split}
&\triangledown\cdot\mathbf{E}_{int}=0,\\
&\partial_{t}\mathbf{E}_{int}=\triangledown\times\mathbf{B}_{int}-\sigma\,(\mathbf{E}_{int}+\mathbf{E}_{ext})-\sigma_{\chi}(\mathbf{B}_{int}+\mathbf{B}_{ext}),\\
&\triangledown\cdot\mathbf{B}_{int}=0,\\
&\partial_{t}\mathbf{B}_{int}=-\triangledown\times\mathbf{E}_{int}.
\end{split}
\end{align}
We numerically solve this equation set to obtain the internal electric and magnetic fields in the medium at any time, and then resulting the dynamical electric and magnetic field in heavy-ion collisions is obtained by adding the external and internal parts. 

For numerical stability for the conductivity ranging from $0$ to $\infty$ when solving Eq.~\eqref{eq:meint}, we chose the Yee's algorithm~\cite{Yee:1966} which belongs to the category of Leapfrog algorithms. In Yee's algorithm, the computed fields $\mathbf{E}$ and $\mathbf{B}$ are staggered by half a step in space-time with respect to each other. More details about the algorithm of Eq.~\eqref{eq:meint} is presented in Appendix~\ref{ap:algorithm-Minkowski}. The code package of this section is publicly available at \url{https://github.com/brangja/EB-in-HIC.git}.

\begin{figure}[!hbt]\centering
    \includegraphics[width=0.45\textwidth]{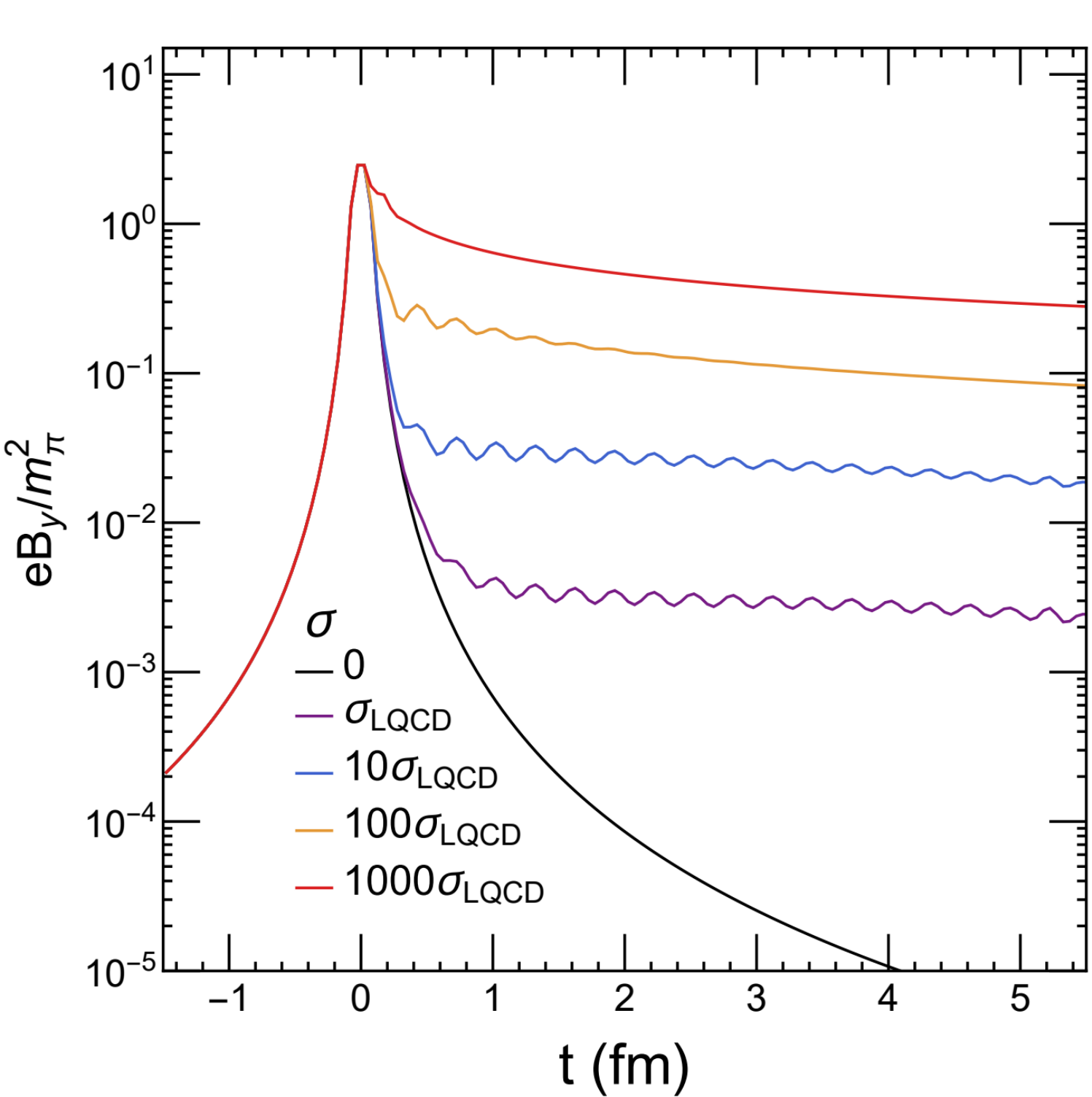}
	\caption{The dynamical magnetic field in the static medium.
	\label{By-static}}
\end{figure}
\subsection{numerical results}
In this subsection, we present the numerical results with the aforementioned numerical method. The simulation is performed using initial condition of the electromagnetic field provided by event-averaged MC-Glauber simulation for Au+Au collisions at RHIC energy $\sqrt{s_\mathrm{NN}}=200$~GeV and impact parameter $b=6$~fm. The velocity and Lorentz factor for both target and projectile can be estimated by $v^{2}=1-\gamma^{-2}$ and $\gamma=\sqrt{s_\mathrm{NN}}/2m_p$. To explore the effect of conductivity, we parametrize the the electric conductivity as the result of hot QCD medium is obtained in the lattice calculations~\cite{Ding:2010ga} scaled by a factor ($\lambda$), 
\begin{align}
&\sigma=\lambda~ \sigma_\mathrm{LQCD} 
    = 5.8\,\lambda~\text{MeV}.
\end{align} 
To test the stability of our program and investigate influence of the conductivity on the evolution of the magnetic field, the parameter is chosen as $\lambda=1$, $10$, $100$, and $1000$. The chiral conductivity is~\cite{Kharzeev:2009pj}  
\begin{align}
	&\sigma_{\chi}=\bigg(\frac{e^{2}}{2\pi^{2}}N_{c}\sum_{f}q^{2}_{f}\bigg)\mu_{5},
\end{align}
where $\mu_5$ is the chiral chemical potential. Based on these inputs and the aforementioned method, we numerically solve the time evolution of the magnetic field in the origin $\mathbf{x}=0$ as a function of the electric conductivity. Results are shown in Fig.~\ref{By-static}, which are qualitatively consistent with McLerran and Skokov~\cite{McLerran:2013hla}\footnote{We speculate the difference in qualitative value to be due to the fact that a different colliding system was consider in~\cite{McLerran:2013hla}. See also~\cite{Zakharov:2014dia}.}. In our calculation, we took two values of the chiral chemical potential, an optimistic limit that $\mu_{5}=1$~GeV and a pessimistic limit that $\mu_{5}=0$. We find the difference between these two cases to be negligible.   

\begin{figure}[!hbt]\centering
	\includegraphics[width=0.45\textwidth]{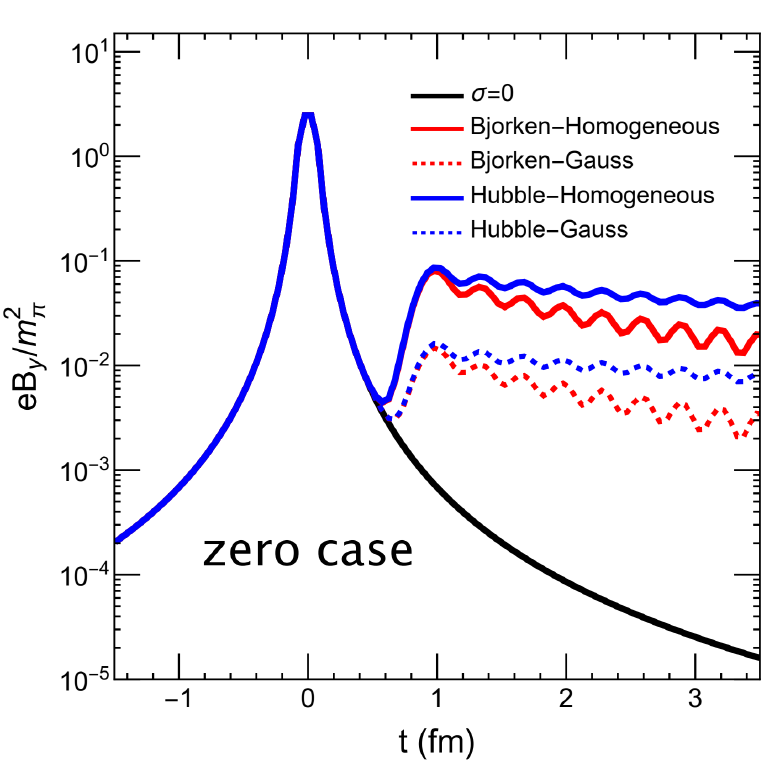}
	\includegraphics[width=0.45\textwidth]{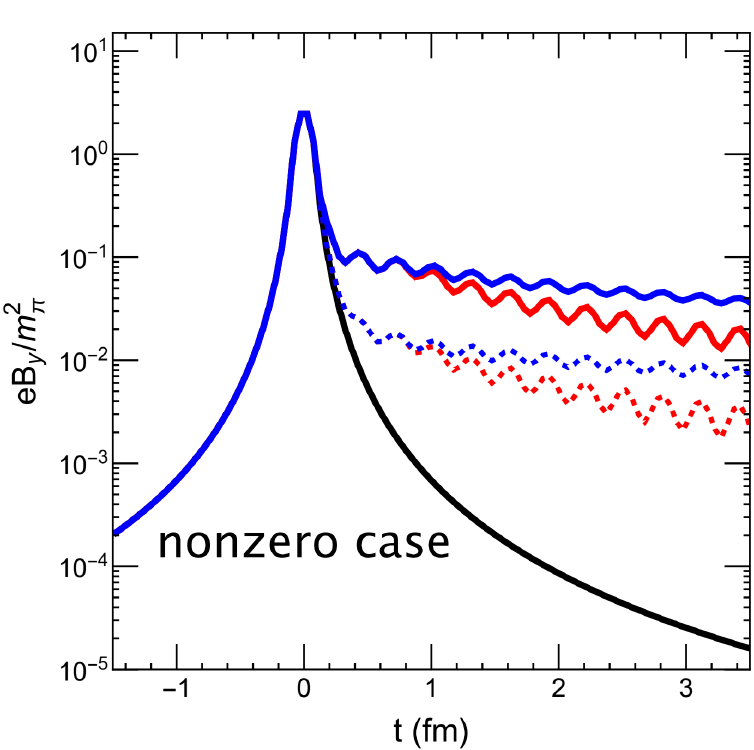}
	\caption{Time dependence of the magnetic field along the out-of-plane direction. Upper and lower panels compare the effect of pre-equilibrium conductivity~\protect{\eqref{eq:test_preeq}}. In each plot, red and blue lines compare the temporal distribution~\protect{\eqref{eq:test_time}}, whereas solid and dashed lines compare the spatial distribution~\protect{\eqref{eq:test_spatial}}. \label{photo-By-sigma}}
\end{figure}

The rapidly expanding conducting medium created in heavy-ion collisions is highly inhomogeneous. A realistic simulation requires a space-time dependent electric conductivity. 
Meanwhile, quarks are not formed immediately after the initial collisions, and it take finite time for the quark to equilibrate. One needs to pay special attention to the non-trivial time dependence of $\sigma$ in the pre-equilibrium stage.

Herein, we test the influence of different components by choosing a couple of different models in the simulation. To explore the limits without transverse expansion and with the strongest one, we take two the Bjorken and Hubble models to estimate the time-dependent of electric conductivity, $\sigma(t)$, i.e.,
\begin{align}
\begin{split}
	\text{Bjorken:}\qquad& \sigma(t)=\sigma_{c}\frac{T(t)}{T_{c}}=\sigma_{c}\frac{T_{0}}{T_{c}}\left(\frac{t_{0}}{t}\right),\\
	\text{Hubble:}\qquad& \sigma(t)=\sigma_{c}\frac{T(t)}{T_{c}}=\sigma_{c}\frac{T_{0}}{T_{c}}\left(\frac{t_{0}}{t}\right)^{\frac{1}{3}}.
\end{split}\label{eq:test_time}
\end{align}
In our calculation, we take $T_{0}=3\,T_{c}$ and $\sigma_{c}=10\,\sigma_\mathrm{LQCD}$.
The spacial dependent is then modeled by assuming homogeneous or Gaussian profile,
\begin{align}
\begin{split}
	\text{Homogeneous:} \qquad& \sigma(t,\mathbf{x})=\sigma(t),\\
	\text{Gaussian:} \qquad& \sigma(t,\mathbf{x})=\sigma(t)e^{-\frac{x^2}{R_{x}^{2}}-\frac{y^2}{R_{y}^{2}}-\frac{z^2}{R_{z}^{2}}},
\end{split}\label{eq:test_spatial}
\end{align} 
where the corresponding parameters at the Gaussian model are $R_x=R_A-b/2$, $R_y=\sqrt{R_A^{2}-b^2/4}$, $R_z=3$~fm, $R_A=6.38$~fm the radius of Au, $b=6$~fm the impact parameter. Finally, the effect of quark formation time is estimated by taking two limits --- quarks at time $0\leq t\leq t_0$ are not created at all, or rapidly created to its density at $t_0$. Correspondingly, the conductivity is parametrized as
\begin{align}
\begin{split}
\text{zero:}&
    \quad 
    \left\{\begin{array}{ll}
    \sigma(t) = 0, & t \leq t_0, \\
    \sigma(t) \neq 0, & t_0 < t;
    \end{array}\right.\\~\\
\text{non-zero:}&
    \quad 
    \left\{\begin{array}{ll}
    \sigma(t)  = 0, & t \leq 0, \\
    \sigma(t) = \sigma(t_0), & 0 < t \leq t_0, \\
    \sigma(t) \neq 0, & t_0 < t.
    \end{array}\right.
\end{split}\label{eq:test_preeq}
\end{align}
Here the $t_0=0.4$ fm is the start time of the hydro stage which is assumed as global equilibrium. The picture of these two cases is that the system is under the pre-equilibrium stage at time interval $0<t<t_0$, and then the system is close to the equilibrium stage after time $t=t_0$. the difference is on the conductivity at the pre-equilibrium stage.  
Corresponding numerical results are showed in Fig.~\ref{photo-By-sigma}. It shows that the $B(t)$ is very sensitive to the effective conductivity of the pre-equilibrium stage ($0 < t < t_0$) and the hydro stage ($t \ge t_0$).
We note that in Ref.~\cite{Stewart:2021mjz}, the authors solve the Maxwell equation for a $\sigma(t)$ that is similar to our ``zero'' scenario, and in Fig.~\ref{photo-By-sigma}, we observe similar behavior for $B(t)$.

\section{Dynamical magnetic fields in the non-static QGP} \label{sec-2}

\subsection{the Maxwell equation in Milne space}
It is convenient to work on the Milne space for investigating the dynamic evolution of the electromagnetic field in the rapidly expanding QGP. The Maxwell equation can be expressed as 
\begin{align}
	&\hat{D}_{\mu}F_{M}^{\mu\nu}=J^{\nu},\label{eq:maxm1}\\
	&\hat{D}_{\mu}\widetilde{F}_{M}^{\mu\nu}=0,\label{eq:maxm2}
\end{align} 
where the electromagnetic field tensor is marked with the subscript $M$ to refer to Milne coordinate. The covariant derivative $\hat{D}_{\mu}$ acting on a tensor is expressed as $\hat{D}_{\mu}t^{\nu\rho}=\partial_{\mu}t^{\nu\rho}+\Gamma^{\nu}_{\lambda\mu}t^{\lambda\rho}+\Gamma^{\rho}_{\lambda\mu}t^{\nu\lambda}$, with affine connections $\Gamma^{\rho}_{\mu\nu} = (1/2) g^{\rho\sigma}(\pt_{\nu}g_{\sigma\mu}+\pt_{\mu}g_{\sigma\nu}-\pt_{\sigma}g_{\mu\nu})$. We adopt the metric convention to be $g_{\mu\nu} = \mathrm{diag}(1,-1,-1,-\tau^{2})$. The dual tensor is $\widetilde{F}_{M}^{\mu\nu}=(1/2)\epsilon^{\mu\nu\rho\sigma}F^{M}_{\rho\sigma}$. Herein, the Levi--Civita tensor $\epsilon^{\mu\nu\rho\sigma}$ and the electromagnetic tensor are different from the case of Minkowski coordinate, and their explicit forms can be found in Appendix~\ref{ap:levi-civita}. The currents are composed of normal currents, diffusion current, Ohm's law, and CME current as the following,   
\begin{align}\begin{split}\label{eq:j}
		&J^{\mu}=J^{\mu}_{in}+J^{\mu}_{s},\\
		&J^{\mu}_{in}=n\,u^{\mu}+d^{\mu}+\sigma\,F^{\mu\nu}_{M}u_{\nu}+\sigma_{\chi}\widetilde{F}^{\mu\nu}_{M}u_{\nu}.	
\end{split}\end{align} 
Herein, $J^{\mu}_{in}$ denotes the current in the medium, and $J^{\mu}_{s}$ the source contributions from the fast moving charged particles in heavy-ion collisions, $n$ is the charge number density and $d^\mu$ the diffusive current. 
We further denote electric and magnetic fields in the Milne space as
\begin{align}
	&\widetilde{E}^{i}=F_{M}^{i0},\qquad \widetilde{B}^{i}=\widetilde{F}_{M}^{i0},
\end{align}
with $i$ being $x$, $y$, or $\eta$.
The current in the medium can be further simplified as follows,
\begin{align}\begin{split}
J^{\mu}_{in}=\;&
	(J^{\tau},J^{x},J^{y},J^{\eta}), \\
J^{\tau}=\;&
	n\,u^{\tau}+d^{\tau}+\sigma\left(\widetilde{E}^{x}u^{x}+\widetilde{E}^{y}u^{y}+\tau^{2}\widetilde{E}^{\eta}u^{\eta} \right)
\\&
	+\sigma_{\chi}\left(\widetilde{B}^{x}u^{x}+\widetilde{B}^{y}u^{y}+\tau^{2}\widetilde{B}^{\eta}u^{\eta} \right),\\
J^{x}=\;&
	n\,u^{x}+d^{x}+\sigma\left(\widetilde{E}^{x}u^{\tau}+\tau\widetilde{B}^{\eta}u^{y}-\tau\widetilde{B}^{y}u^{\eta} \right)
\\&
	+\sigma_{\chi}\left(\widetilde{B}^{x}u^{\tau}-\tau\widetilde{E}^{\eta}u^{y}+\tau\widetilde{E}^{y}u^{\eta} \right),\\
J^{y}=\;&
	n\,u^{y}+d^{y}+\sigma\left(\widetilde{E}^{y}u^{\tau}-\tau\widetilde{B}^{\eta}u^{x}+\tau\widetilde{B}^{x}u^{\eta} \right)
\\&
	+\sigma_{\chi}\left(\widetilde{B}^{y}u^{\tau}+\tau\widetilde{E}^{\eta}u^{x}-\tau\widetilde{E}^{x}u^{\eta} \right),\\
J^{\eta}=\;&
    n\,u^{\eta}+d^{\eta}+\sigma\left(\widetilde{E}^{\eta}u^{\tau}+\frac{\widetilde{B}^{y}}{\tau}u^{x}-\frac{\widetilde{B}^{x}}{\tau}u^{y} \right)
\\&
    +\sigma_{\chi}\left(\widetilde{B}^{\eta}u^{\tau}-\frac{\widetilde{E}^{y}}{\tau}u^{x}+\frac{\widetilde{E}^{x}}{\tau}u^{y} \right).	
\end{split}\end{align}

In order to facilitate the subsequent numerical calculations, let us further simplify the above Maxwell equations. From Eq.~\eqref{eq:maxm1} and~\eqref{eq:maxm2}, one can get the evolution equations of the electric and magnetic fields as,
\begin{align}
\begin{split}
	&\partial_{x}\widetilde{E}^{x}+\partial_{y}\widetilde{E}^{y}+\partial_{\eta}\widetilde{E}^{\eta}=J^{\tau},\\
	&\partial_{\tau}(\tau\,\widetilde{E}^{x})=\partial_{y}(\tau^{2}\widetilde{B}^{\eta})-\partial_{\eta}\widetilde{B}^{y}-\tau\,J^{x},\\
	&\partial_{\tau}(\tau\,\widetilde{E}^{y})=-\partial_{x}(\tau^{2}\widetilde{B}^{\eta})+\partial_{\eta}\widetilde{B}^{x}-\tau\,J^{y},\\
	&\partial_{\tau}(\tau\,\widetilde{E}^{\eta})=\partial_{x}\widetilde{B}^{y}-\partial_{y}\widetilde{B}^{x}-\tau\,J^{\eta}.
\end{split}\label{eq:ME-Milne-E}\\~\nonumber\\
\begin{split}
		&\partial_{x}\widetilde{B}^{x}+\partial_{y}\widetilde{B}^{y}+\partial_{\eta}\widetilde{B}^{\eta}=0,\\
		&\partial_{\tau}(\tau\,\widetilde{B}^{x})=-\partial_{y}(\tau^{2}\widetilde{E}^{\eta})+\partial_{\eta}\widetilde{E}^{y},\\
		&\partial_{\tau}(\tau\,\widetilde{B}^{y})=\partial_{x}(\tau^{2}\widetilde{E}^{\eta})-\partial_{\eta}\widetilde{E}^{x},\\
		&\partial_{\tau}(\tau\,\widetilde{B}^{\eta})=-\partial_{x}\widetilde{E}^{y}+\partial_{y}\widetilde{E}^{x}.
\end{split}\label{eq:ME-Milne-B}
\end{align}

Now we can carry out the simulations with above equations~\eqref{eq:ME-Milne-E} and~\eqref{eq:ME-Milne-B}, according to the aforementioned numerical method in Section \ref{sec:NME}. In final numerical results, we will compute the electric and magnetic field in Minkowski coordinates, which are Lorentz transformation of electric and magnetic field in Milne coordinate by the following,
\begin{align}\begin{split}
E^{x}=\;&\cosh\eta\,\widetilde{E}^{x}+\sinh\eta\,\widetilde{B}^{y}, \\
E^{y}=\;&\cosh\eta\,\widetilde{E}^{y}-\sinh\eta\,\widetilde{B}^{x},\\
E^{z}=\;&\tau\widetilde{E}^{\eta},\\
B^{x}=\;&\cosh\eta\,\widetilde{B}^{x}-\sinh\eta\,\widetilde{E}^{y},\\
B^{y}=\;&\cosh\eta\,\widetilde{B}^{y}+\sinh\eta\,\widetilde{E}^{x},\\
B^{z}=\;&\tau\widetilde{B}^{\eta}.
\end{split}\end{align}

\subsection{Electric Conductivity}	

The electric conductivity of QGP remains to be an open question. There are many works focusing on the electric conductivity in the hydro stage of QGP with theoretical calculations and simulations, but significantly different results are obtained. Here we briefly outline some of them in the following for further estimation of the reasonable region. First, results from different lattice QCD calculations can be different by an order of magnitude. They are listed as follows,
\begin{itemize}
    \item[\cite{Gupta:2003zh}] $\frac{\sigma}{T}|_{1.5<T/T_{c}<3}=7C_{em}\approx 0.428$,
    \item[\cite{Aarts:2007wj}] $\frac{\sigma}{T}|_{T/T_c\approx 1.5}=(0.4\pm 0.1)C_{em}=0.0245\pm 0.006$,
    \item[\cite{Ding:2010ga,Ding:2016hua}] 
        $\frac{\sigma}{T}|_{1.1T_{c}} \!=\! (0.201 \!\sim\! 0.703)C_{em}\!\approx\!(1.23 \!\sim\! 4.30)\times10^{-2}$,\\
        $\frac{\sigma}{T}|_{1.3T_{c}} \!=\! (0.203 \!\sim\! 0.388)C_{em}\!\approx\!(1.24 \!\sim\! 2.37)\times10^{-2}$,\\
        $\frac{\sigma}{T}|_{1.5T_{c}} \!=\! (0.218 \!\sim\! 0.413)C_{em}\!\approx\!(1.33 \!\sim\! 2.52)\times10^{-2}$,
\end{itemize}
where the factor $C_{\text{em}}=\sum_{f}e^{2}_f\approx0.06115$ for three-flavor case, $e_f$ is the charge of quark with flavor $f$. 

Additionally, results from different theoretical calculations are also different. The hard thermal loop(HTL) calculation up to leading-log for high temperature QGP produces that $\sigma/T=11.8687e\frac{\Tr_{f}(Q_{e}Q_{V})}{g^{4}\ln(1/g)}=146.33$($26.12$) for $\alpha_{s}=0.01$($0.05$)~\cite{Jiang:2014ura}, where $Q_{e}=(2/3, -1/3, -1/3)$ for $(u, d, s)$ and the numerical results is for $Q_{V}=Q_e $. Meanwhile, the leading order perturbative QCD calculation gives that $\sigma/T\approx 5.98$~\cite{Arnold:2003zc}, whereas the dilute instanton-liquid model gives that $\sigma/T\approx\left(0.46\sim 1.39\right)C_{em}\approx\left(0.0281\sim 0.0850\right)$~\cite{Nam:2012sg}. The transport model with relaxation time gives us an analytical representation, (see e.g.,~\cite{Thakur:2017hfc,Kadam:2017iaz,Das:2019wjg,Thakur:2019bnf,Das:2019ppb})
\begin{align}
\begin{split}
	\sigma=\;& \sum_{f,\pm} \frac{g_{f}q^{2}_{f}\tau_{q}}{6\pi^2 T}
	\int\frac{k^2 dk }{E^{2}_{k,f}}
	\frac{e^{-\frac{E_{k,f}\pm\mu}{T}}}{(e^{-\frac{E_{k,f}\pm\mu}{T}}+1)^2},
\end{split}
\end{align}
where $+$($-$) sign is taken for fermion(antifermion). One can estimate $\sigma/T=0.007\sim0.026$ for $T=T_{c}-5T_{c}$ with zero chemical potential
\footnote{In this estimation, the relaxation time is $\tau_{q}=\frac{1}{5.1T\alpha_{s}^{2}\ln(\alpha_{s}^{-1})\left(1+0.12\left(2N_{f}+1\right)\right)}$, $g_f=2\times3$, the mass taking the effective mass composed of the bare mass and thermal mass, and the coupling constant $\alpha_{s}(T)=\frac{6\pi}{(33-2N_{f})\ln(T/\Lambda_{T})}\left(1-\frac{3(153-19N_{f})}{(33-2N_{f})^{2}}\frac{\ln(2\ln(T/\Lambda_{T}))}{\ln(T/\Lambda_{T})}\right)$, with $\Lambda_T = 200$~MeV.}.
Furthermore, the parton-hadron-string dynamics (PHSD) transport~\cite{Cassing:2013iz} approach finds that $\sigma/T\approx 0.0009+0.015 (T-T_{c} )/T_{c}$.Finally, the microscopic relativistic transport model Boltzmann Approach to Multi-Parton Scatterings(BAMPS) simulation~\cite{Greif:2014oia} obtained $\sigma/T \approx (0.05\sim0.2)$.

Based on above summary, we choose $\sigma/T=0.1$ as a relatively reasonable value in our numerical simulation. 
We also take $\sigma/T=100$ to explore the medium response in the large conductivity limit.

\subsection{Numerical results}	
In this subsection, we present the simulate results of the electromagnetic field evolution in the non-static QGP. As mentioned before, the evolution of the medium formed in the relativistic heavy-ion collision, as well as that of the electromagnetic field, consists of three stages --- initial($\tau<0.1$~fm), pre-equilibrium($0.1\leq\tau<0.4$~fm), and hydro($\tau\geq0.4$~fm) stage. Our simulation will be arranged accordingly.  The initial condition of the electromagnetic field is generated by two heavy nuclei moving toward each other. Then we solve the Maxwell's equations~\eqref{eq:ME-Milne-E} and~\eqref{eq:ME-Milne-B} and simulate the electromagnetic field evolution in the pre-equilibrium and stage. In the pre-equilibrium stage, the QGP is assumed to expand as a Bjorken flow, and we explore three different models for the time dependence of the electrical conductivity. Whereas in the hydro stage, we take hydro background from Bjorken flow, Gubser flow, and realistic hydro profile from the MUSIC package~\cite{Gale:2013da,Schenke:2010rr,Schenke:2010nt,McDonald:2016vlt}. the corresponding temperature and fluid velocity are produced by these three models. (See the Appendix~\ref{app:t-v} for analytical forms of temperature and fluid velocity in the Bjorken and Gubser solutions.) These temperature and fluid velocity profile and then read into the program and provide the background field to solve Maxwell's equations~\eqref{eq:ME-Milne-E} and~\eqref{eq:ME-Milne-B}. 
In what follows, we will focus on the dynamical magnetic field at the center of the fireball [$\mathbf{x}=(0, 0, 0)$]\footnote{It is worth noting that we can generate the dynamic magnetic and electric fields at any given coordinate and the spatial distribution of electromagnetic fields at any time under our framework.}. Herein, we focus on the case that the net number density ($n$) and the diffusive current density ($d^{\mu}$) are set to zero, and their influence on the dynamic evolution of the magnetic field will be studied in our future work.

\begin{figure}[!hbt]\centering
	\includegraphics[width=0.45\textwidth]{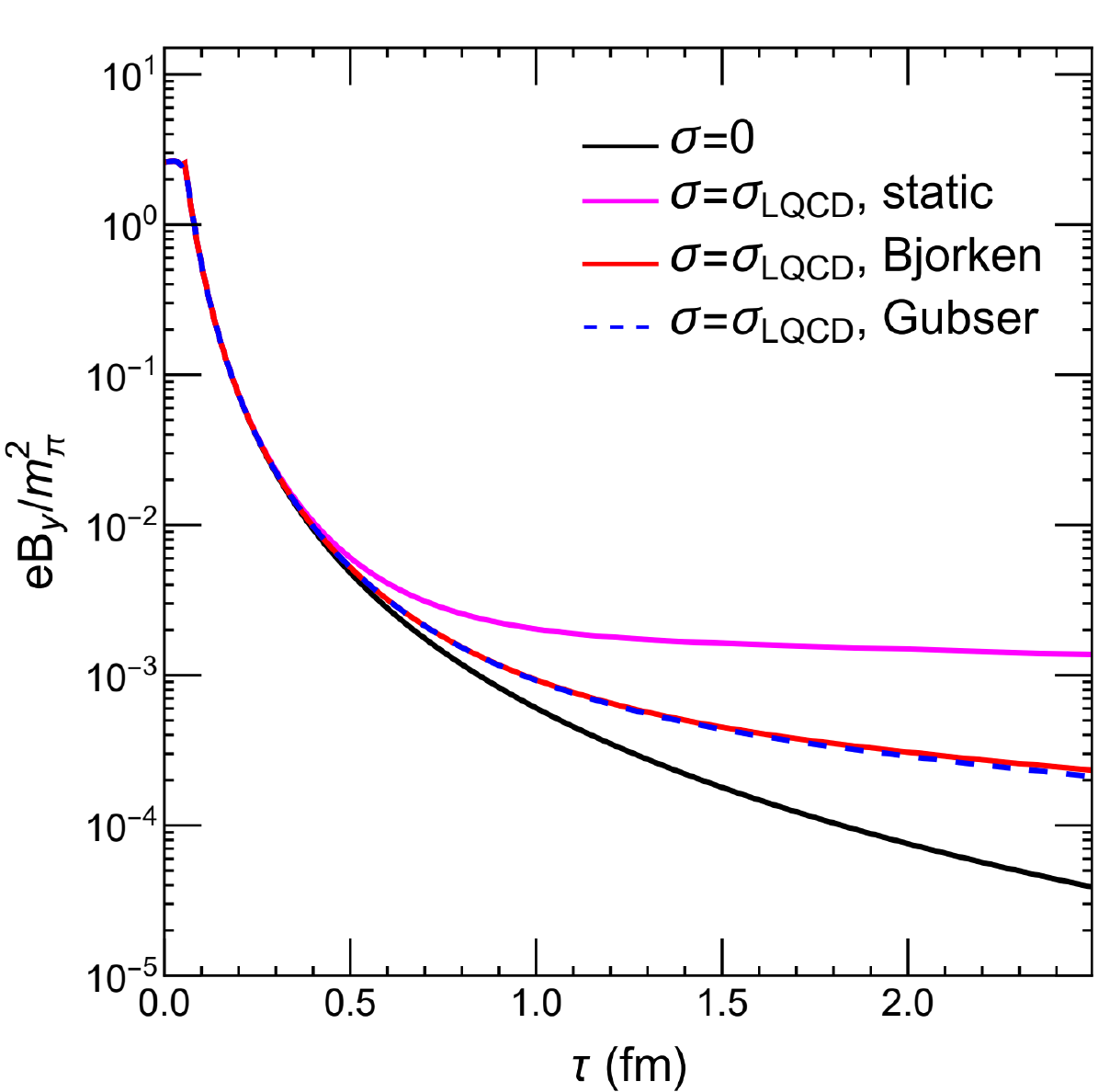}
    \caption{The suppression effect of the longitudinal expansion of the medium. \label{compare-expand-medium}}
\end{figure}
\subsubsection{suppression effect due to the longitudinal expansion of the medium}
We first study the evolution of magnetic field along the out-of-plane direction ($e\,B_y$) with three different backgrounds --- a static medium, Bjorken flow, and Gubser flow. We take identical electric conductivity in these three different cases, i.e., $\sigma=0$ for the initial stage, and $\sigma=\sigma_{\mathrm{LQCD}}$ for the pre-equilibrium and hydro stage. Results are shown in Fig.~\ref{compare-expand-medium}. We observe a suppression effect for the longitudinal expansion of the medium. The dynamical magnetic field was depressed in the Bjorken expansion and the Gubser expansion compared to the static case.

\begin{figure}[!hbt]\centering
    \includegraphics[width=0.45\textwidth]{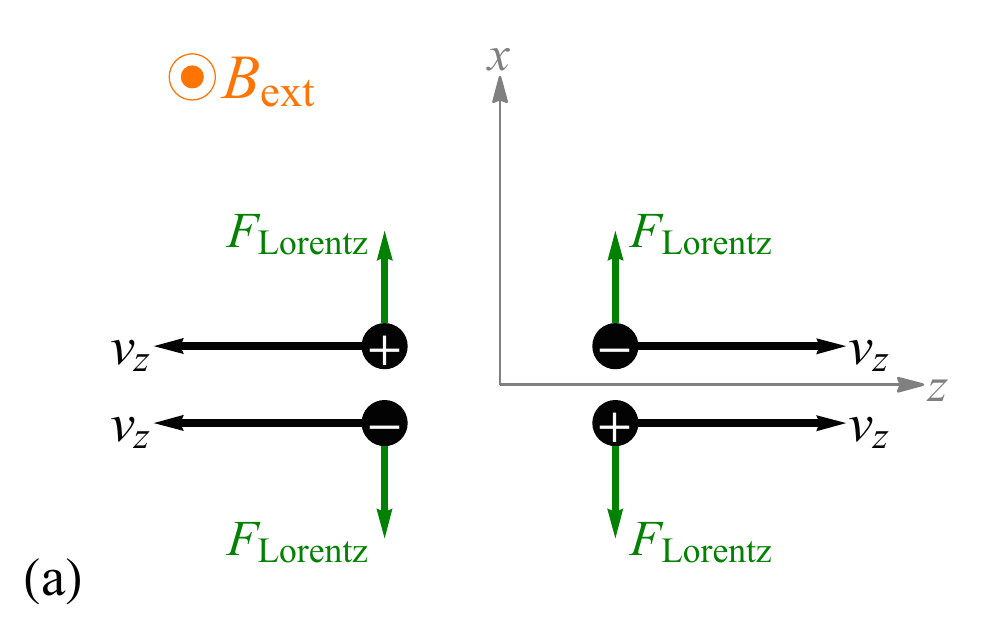}
    \includegraphics[width=0.45\textwidth]{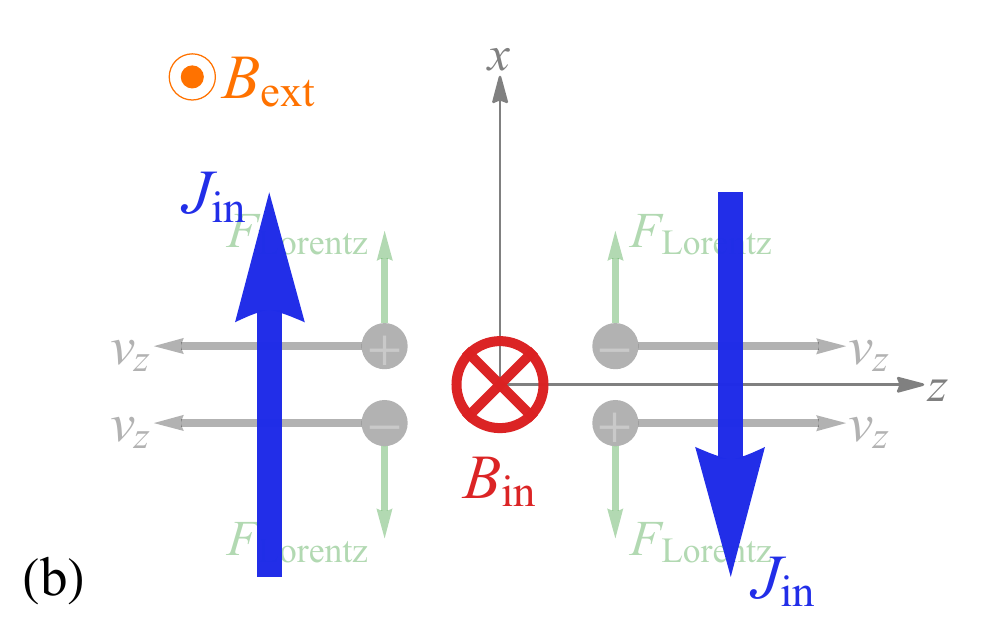}
	\caption{Illustraction of the suppression effect. (a) Lorentz force acting on in-medium charged particles due to the external magnetic field. (b) Induced electric currents due to the Lorentz force and their resulting induced magnetic field. \label{suppression-effect}}
\end{figure}

The reason of the suppression effect is illustrated in Fig.~\ref{suppression-effect}  and explained as follows. At the presence of a external magnetic field along the $y$-axis, in-medium particles with positive(negative) at forward rapidity experience a Lorentz pointing at the negative(positive) $x$-direction, and vice versa for particles at backward rapidity. Collective motion of the charged particles, due to the Lorentz force, induces a clockwise circular current. It generates an induced magnetic field in the negative $y$-direction and thereby weakens the external magnetic field.

\begin{figure*}[!hbt]\centering
	\includegraphics[width=0.31\textwidth]{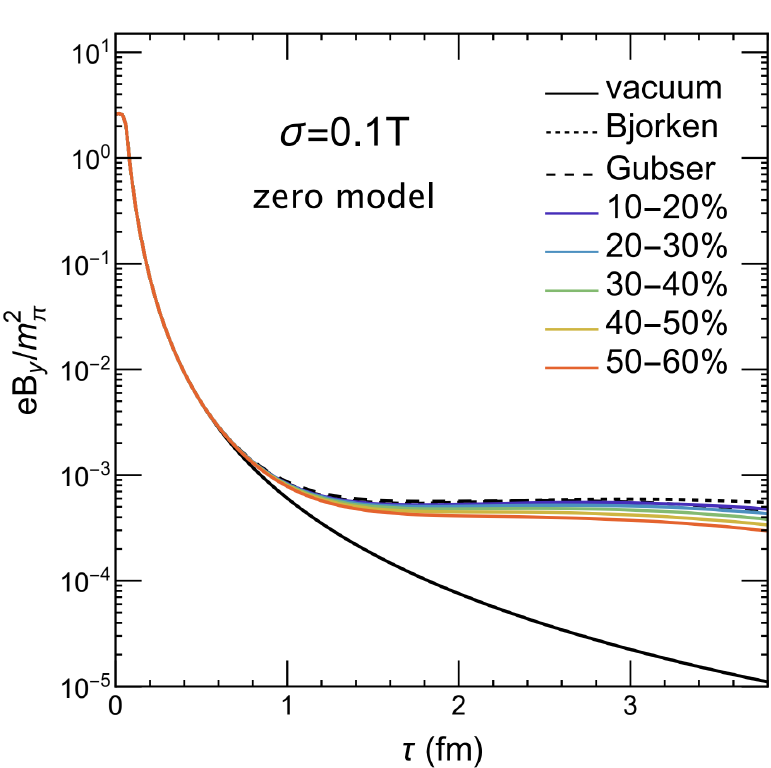}
	\includegraphics[width=0.3\textwidth]{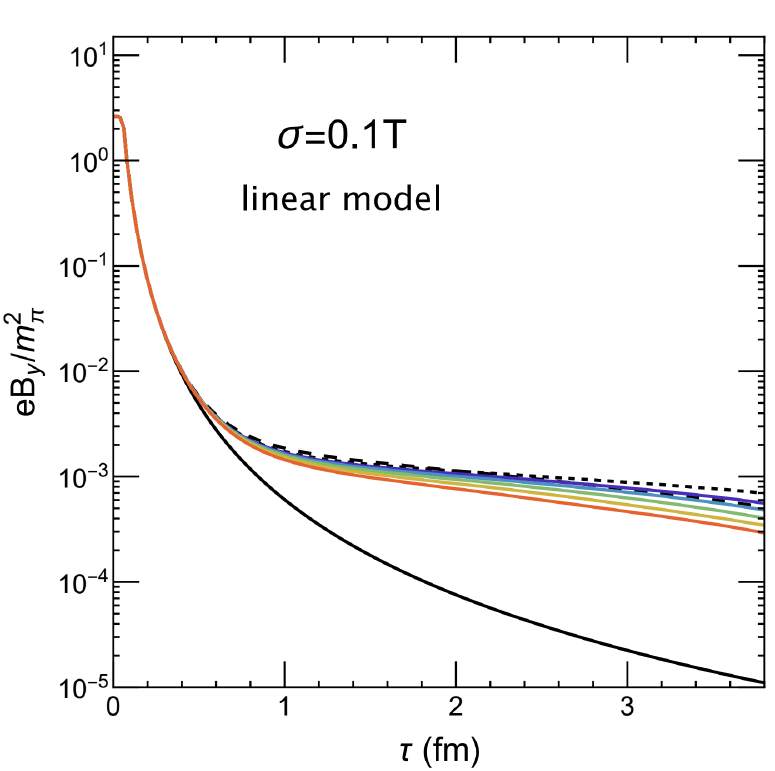}
	\includegraphics[width=0.3\textwidth]{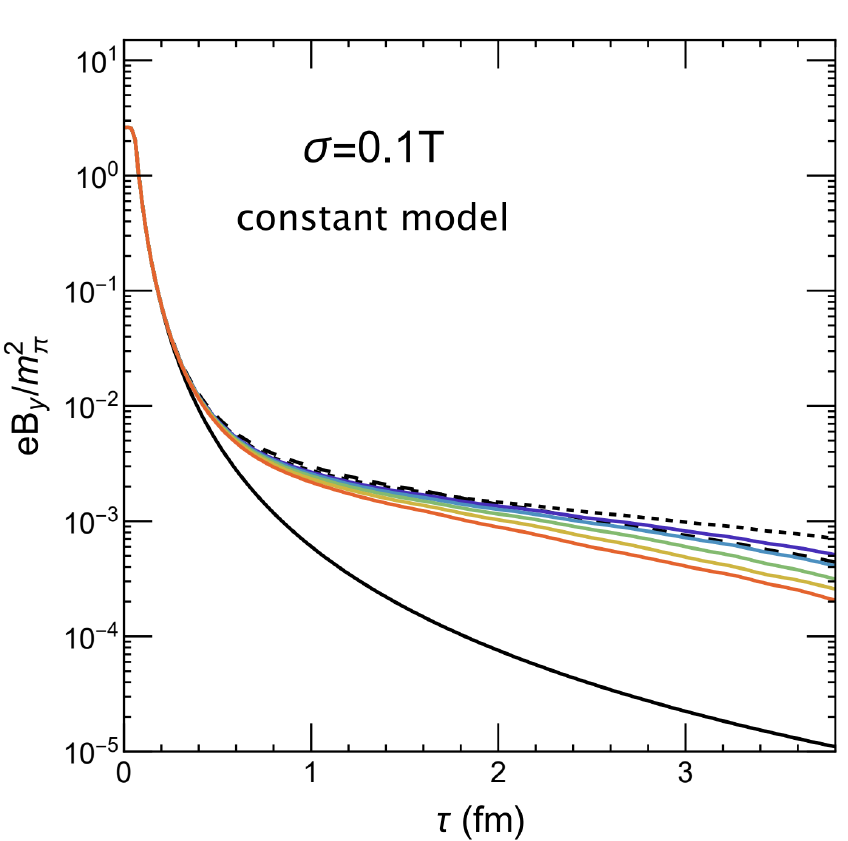}
\caption{Proper dependence of magnetic field along the out-of-plane direction at the center ($x=0,y=0,z=0$) and with conductivity $\sigma=0.1\,T$ in the hydro stage. From left to right correspond to zero, linear, and constant models for the pre-equilibrium stage. Black solid curves represent the vacuum value, whereas dotted(dashed) curves take Bjorken(Gubser) flow as the background. Colored solid curves from purple to red are respectively for realistic hydro background in the $10-20\%$ to $50-60\%$ centrality classes.
		\label{sigma-0.1T}}
	\includegraphics[width=0.31\textwidth]{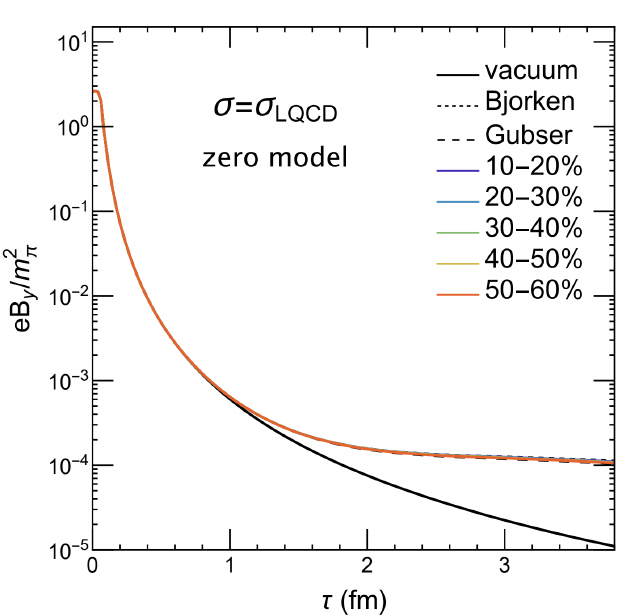}
	\includegraphics[width=0.3\textwidth]{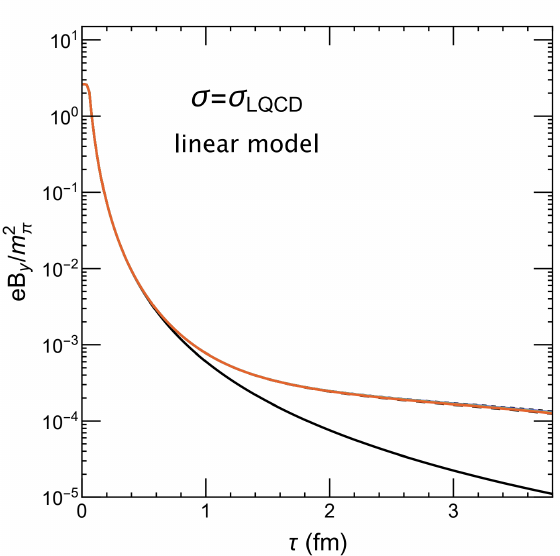}
	\includegraphics[width=0.3\textwidth]{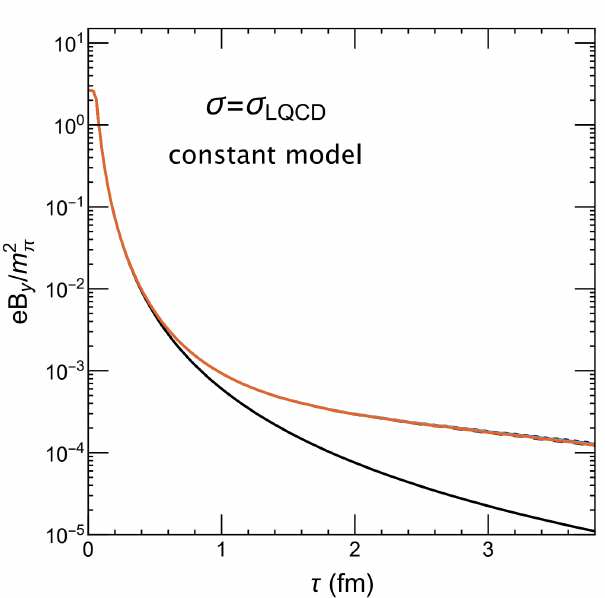}
	\caption{Same as \protect{Fig.~\ref{sigma-0.1T}} but with hydro constant conductivity $\sigma=\sigma_\mathrm{LQCD}$.\label{sigmaLQCD}}
	\includegraphics[width=0.31\textwidth]{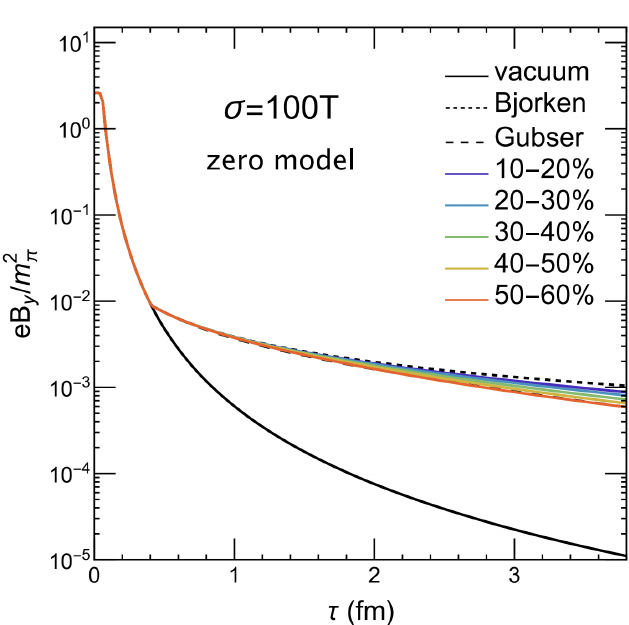}
	\includegraphics[width=0.3\textwidth]{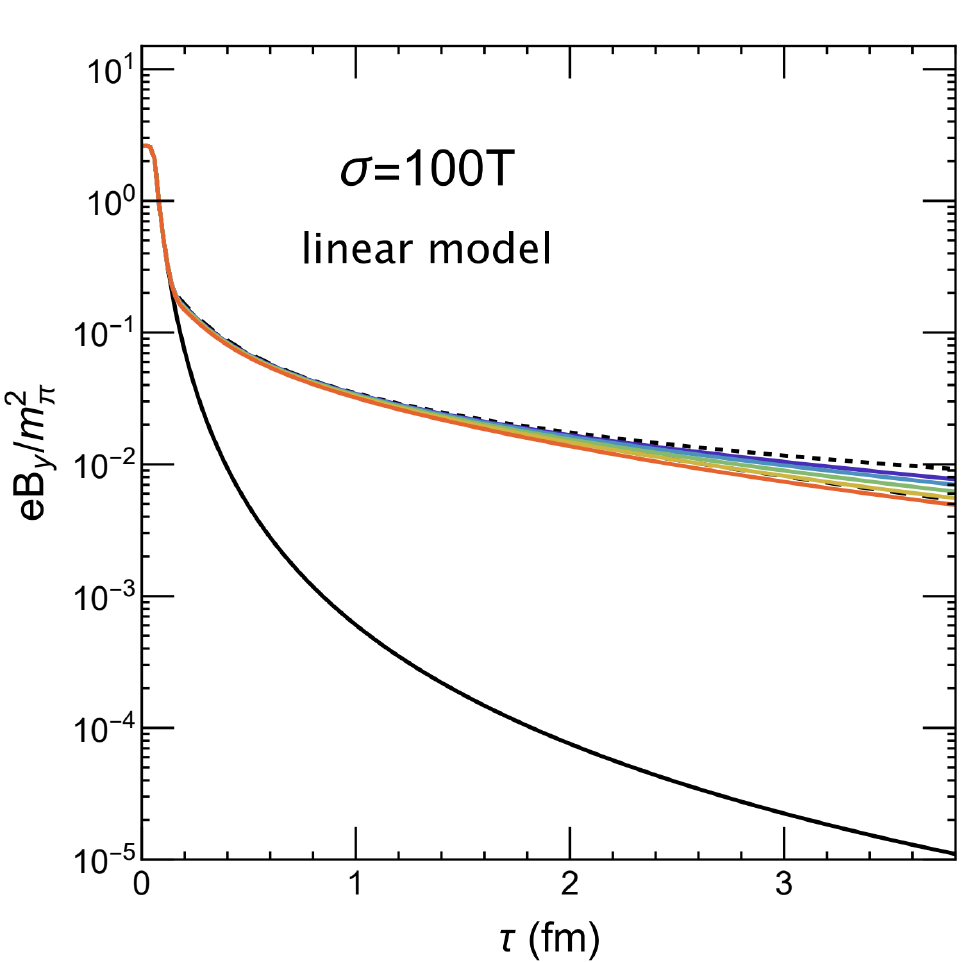}
	\includegraphics[width=0.3\textwidth]{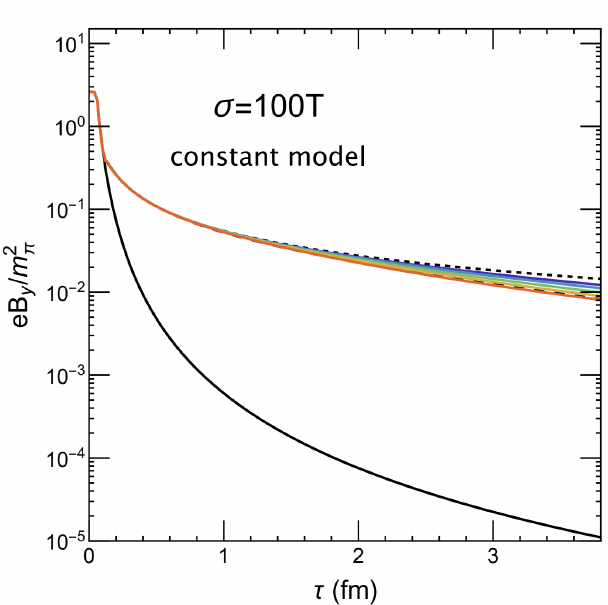}
\caption{Same as \protect{Fig.~\ref{sigma-0.1T}} but with hydro conductivity $\sigma=100\,T$.\label{sigma-100T}}
\end{figure*}

\begin{table*}\centering
\begin{tabular}{| m{3cm} | m{1.5cm}| m{1.5cm} | m{1.5cm} | m{1.5cm}| m{1.5cm} | m{1.5cm}| m{1.5cm} |m{1.5cm} | }
			\hline  & $a_0$ & $b_0$ & $b_1$ & $b_2$ & $c_0$ & $c_1$ & $c_2$ & $c_3$\\
			\hline zero model &-8.848 &-2.193 & 0.412&-0.207&0.0 &13.074 &-11.66 &19.679\\
			\hline constant model &-21.693 &-28.039 &-148.773 &-16.302 &0. &159.619 &6.285 &4.948\\
			\hline linear model &-10.177 &-7.549 &0.183 &-0.299 &0.351 &36.926 &-16.311 &3.285\\
			\hline
\end{tabular}
\caption{The fit parameters of the fitted magnetic field Eq.~\eqref{eq:fit} for electric conductivity $\sigma=0.1\,T$ in hydro stage with different conductivity model in pre-equilbrium stage.\label{table-0.1T}}
\end{table*}
\begin{table*}\centering
\begin{tabular}{| m{3cm} | m{1.5cm}| m{1.5cm} | m{1.5cm} | m{1.5cm}| m{1.5cm} | m{1.5cm}| m{1.5cm} |m{1.5cm} | }
	\hline  & $a_0$ & $b_0$ & $b_1$ & $b_2$ & $c_0$ & $c_1$ & $c_2$ & $c_3$\\
	\hline zero model &-9.924 &-2.145 &-8.568 &3.602 &0. &14.76 &12.063 &14.417\\
	\hline constant model &-10.871 &-1.786 &-5.669 &-7.969 &0.0 &10.273 &16.761 &20.953\\
	\hline linear model &-10.298 &-2.785 &-6.329 &-1.854 &0.0 &16.528 &10.082 &18.5673\\
	\hline
\end{tabular}
\caption{Same as Table.~\protect{\ref{table-0.1T}} but for constant conductivity $\sigma=\sigma_\mathrm{LQCD}$.\label{table-sigma_LQCD}}
\begin{tabular}{| m{3cm} | m{1.5cm}| m{1.5cm} | m{1.5cm} | m{1.5cm}| m{1.5cm} | m{1.5cm}| m{1.5cm} |m{1.5cm} | }
	\hline  & $a_0$ & $b_0$ & $b_1$ & $b_2$ & $c_0$ & $c_1$ & $c_2$ & $c_3$\\
	\hline zero model &-10.676 &-4.081 &-0.93 &-18.58 &0.052 &19.787 &13.497 &14.866\\
	\hline constant model &-5.633 &-1.344 &-4.912 &1.406 &0. &12.345 &-0.039 &0.677\\
	\hline linear model &-7.311 &-1.248 &0.483 &-12.433 &0. &8.03 &-0.223 &16.904\\
	\hline
\end{tabular}
\caption{Same as Table.~\protect{\ref{table-0.1T}} but for conductivity $\sigma=100\,T$.\label{table-100T}}
\end{table*}
\subsubsection{The dynamical magnetic field with realistic hydro background}
Then we move on to realistic hydrodynamic background provided by numerical simulation using the MUSIC package~\cite{Gale:2013da,Schenke:2010rr,Schenke:2010nt,McDonald:2016vlt}. We set $\sigma=0$ at the initial stage, and we explore the medium responses to the magnetic field by choosing three different electric conductivity models in the pre-equilibrium stage and respectively setting the electric conductivity $\sigma=0.1\,T$, $\sigma=\sigma_{\text{LQCD}}$, and $\sigma=100\,T$ in the hydro stage. The three models for the pre-equilibrium stage are taken: (a) zero model that assumes vanishing conductivity, $\sigma(\tau,\mathbf{x})=0$; (b) constant model that assumes constant conductivity which takes the value at the initial time of hydro, $\sigma(\tau,\mathbf{x})=\sigma(\tau=0.4\text{fm},\mathbf{x})$; and (c) linear model that the conductivity that grows linearly from zero to the value at the hydro initial time, $\sigma(\tau,\mathbf{x})=(\tau/0.4\text{fm})\sigma(\tau=0.4\text{fm},\mathbf{x})$.
We will assume that the temperature and velocity of the background flow in the pre-equilibrium stage follow the Bjorken flow.

Initial conditions of both hydro and electromagnetic field are generated by event-averaged MC-Glauber simulation for $\sqrt{s_\mathrm{NN}}=200$~GeV AuAu collisions.
For a more direct comparison of the response of different hydro profiles, we fix the magnetic field to be the one with impact parameter $b=6$~fm, and vary the hydro background from $10-20 \%$ to $50-60 \%$ centrality classes.

Results for $\sigma=0.1T$ in the hydro stage are presented in Fig.~\ref{sigma-0.1T}. The dynamical magnetic field is more sensitive to the electric conductivity model of the pre-equilibrium stage than to the hydro background. The dynamical magnetic fields in Bjorken, Gubser, and realistic backgrounds are almost the same except for the late time region where the megnetic field strength is very small. It means that the evolution of the $y$ component of the magnetic field is dominated by the longitudinal expansion of the QGP, rather than the transverse expansion.
We fit the magnetic field as a function of time with a parameterized function, 
\begin{align}\label{eq:fit}
&e\,B^y=e\,B^y_{\tau=0}\,\mathrm{Exp}\left[a_0\, e^{\frac{b_0+b_1 \tau+b_2 \tau^2}{c_0+c_1 \tau+c_2 \tau^2+c_3 \tau^3}}\right].
\end{align}
Corresponding parameters in Table~\ref{table-0.1T} are fit from the data of the MUSIC hydro background with centrality $40-50\%$.

Then we investigate the evolution of magnetic field with the constant conductivity $\sigma=\sigma_\mathrm{LQCD}$ in the hydro stage. Results are presented in Fig.~\ref{sigmaLQCD} and parameter fit for $40-50\%$ centrality range is listed in Table~\ref{table-sigma_LQCD}. Given the constant conductivity, the influence of transverse expansion become negligible.

Finally, let us explore the response of a highly conductive plasma where the electric conductivity with $1000$ times in the hydro stage, i.e., $\sigma=100\,T$. Results are shown in Fig.~\ref{sigma-100T} and Table.~\ref{table-100T}. Compared with the results in Fig.~\ref{sigma-0.1T}, the late-time strength of the magnetic field is much greater, and it is more explicitly dependent on the choice of the conductivity model in the pre-equilibrium stage.

\section{Conclusion} \label{sec-3}

To conclude, we have developed a framework to numerically simulate the dynamical magnetic fields in heavy ion collisions. This framework has allowed us to investigate the in-medium evolution of space-time-dependent magnetic fields on top of a variety of background medium evolution models for different scenarios of electric conductivities both in the thermal phase and in the pre-equilibrium stage. Our main findings can be summarized as follows. 
\begin{itemize}

\item In the case of a static QGP, previous results assuming constant electric conductivity are reproduced and new results with more realistic space-time-dependent electric conductivity are obtained, demonstrating a robust medium response that extends the lifetime of the magnitude and that is sensitive to the values of the conductivity. 

\item For an expanding QGP, we find a strong influence of the longitudinal expansion which considerably reduces the contributions from medium response and as a result leads to a much smaller magnetic fields as compared with the static case. On the other hand, the inclusion of transverse expansion in addition to the longitudinal expansion only affects the dynamical field evolution rather mildly. 

\item The lifetime of the dynamical magnetic fields is strongly dependent on the medium conductivities in the thermal QGP. Choosing a conductivity value in the range implied by relevant lattice simulations would only lead to a limited medium enhancement of late time field strength. 

\item More importantly, the lifetime is found to be  particularly sensitive to, and mainly determined by, the   non-equilibrium contribution from the early time partonic medium, as demonstrated by comparisons among the three different choices (zero model, constant model and linear model). A considerable pre-hydro effective conductivity could significantly enhance the dynamical field strength.  

\end{itemize}

Clearly, the main ``bottleneck'' for an accurate description of the dynamical magnetic fields is a better estimate of the effective conductivity for the pre-equilibrium stage, which in turn relies on a detailed understanding of the pre-thermal evolution (--- especially that of the quarks and antiquarks).  From a phenomenological perspective, a magnetic field lifetime on the order of $\sim 1~\mathrm{fm}/c$ in $\sqrt{s_\mathrm{NN}}=200$~GeV collisions appears needed for explaining relevant observables of chiral magnetic effect and the $\Lambda/\bar{\Lambda}$ global polarization splitting~\cite{Muller:2018ibh, Guo:2019mgh, Guo:2019joy, Shi:2017cpu, Shi:2019wzi, Xu:2022hql}. According to our findings in this work, such a lifetime would suggest a considerable medium response contribution at the very early stage of the collisions. Whether this scenario could realistically occur will be an important question for future investigation.  

Finally, while this work focuses on $\sqrt{s_\mathrm{NN}}=200$~GeV Au+Au collisions,  our framework can be readily applied for other colliding systems at different collision energies. The simulation code for the static case has been made available at \url{https://github.com/brangja/EB-in-HIC.git} and efforts are underway to make the full dynamical package publicly available in the future. Such a framework will allow quantitative estimates of many interesting observables induced by the dynamical magnetic fields in heavy ion collisions. 

\section*{Acknowledgments}
The authors thank Dmitri E. Kharzeev for very helpful discussions at the early stage of this work. The research of AH is supported by the National Natural Science Foundation of China (NSFC) Grant Nos.12205309. AH and MH are supported by the Strategic Priority Research Program of Chinese Academy
of Sciences Grant No. XDB34030000, the Fundamental Research Funds for the Central Universities, and the National Natural Science Foundation of China (NSFC) Grant Nos. 12235016.
MH is also supported in part by the National Natural Science Foundation of China (NSFC) Grant Nos.11725523, 11735007, and 12221005. DS is grateful to support by Key Laboratory of Quark and Lepton Physics Contracts No.QLPL2022P01.
JL acknowledges partial support by NSF Grant No. PHY-2209183 and DOE-funded Beam Energy Scan Theory (BEST) Collaboration.
SS is supported by the U.S. Department of Energy, Office of
Science, Office of Nuclear Physics, Grants Nos. DE-FG88ER41450 and DE-SC0012704.

\begin{appendix}
\begin{widetext}
\section{The solutions of Eq.(\ref{eq:meex})}\label{ap:EB-moving-ion}\label{app-1}
As mentioned before, we do not need to solve the Eq.(\ref{eq:meex}), it can be represented analytically by a boosted point charge. If we choose the distribution of
protons both in projectile and target according to the Woods-Saxon distribution
with the standard parameters\cite{Alver:2008aq}. 
\begin{align}
&\rho(r)=\rho_{0}\frac{1+w\,(\frac{r}{R})^{2}}{1+e^{\frac{r-R}{a}}}.
\end{align} 
To Au nucleus, $a=0.535$ fm, $R=6.38$ fm, $w=0$. In this work, we will use the Gauss theorem in the rest frame of the target, and then boost the electromagnetic fields to the lab frame with the velocity of the target. In the rest frame of the target,
\begin{align}
&\mathbf{E}_{0}=\frac{e}{4\pi}Q(r_{0})\frac{\mathbf{r}_{0}}{r_{0}^{3}},\qquad Q(r)=4\pi\int^{r}_{0} \rho(x)x^{2}dx,\qquad \mathbf{B}_{0}=0,\qquad \mathbf{r}_{0}=\mathbf{x}-\mathbf{x}^{'}.
\end{align}
Where $\mathbf{x}$ and $\mathbf{x}'$ are the location of the field and nucleus center respectively. After boosting to the lab frame,
\begin{align}
\begin{split}
&E_{x}=\gamma E^{0}_{x},\qquad E_{y}=\gamma E^{0}_{y}, \qquad E_{z}=E^{0}_{z},\\
&B_{x}=-\gamma\,v E^{0}_{y},\qquad B_{y}=\gamma\,v E^{0}_{x}, \qquad B_{z}=0.
\end{split}
\end{align}
The corresponding coordinates represented by the lab frame are
\begin{align*}
r_{0,x}=x-x^{'},\qquad r_{0,y}=y-y^{'},\qquad r_{0,z}=\gamma(z-z^{'}-vt)
\end{align*}

One can also generate the electromagnetic fields of heavy-ion collisions by using the Monte Carlo method like the work of Deng and Huang~\cite{Deng:2012pc}. In this work, we will use the above method for simplicity.

\section{The algorithm of Yee-grid method to Eq.~\eqref{eq:meint}}\label{ap:algorithm-Minkowski}\label{app-2}
The Yee-grid algorithm of Eq.~\eqref{eq:meint} can be easily understood by the following representation,
\begin{align}
\begin{split}
&\frac{dE_x}{dt}\left(t+dt/2,x+dx/2, y, z\right)=\left[\partial_{y}B_{z}-\partial_{z}B_{y}-\sigma\,E_{x}-\sigma_{\chi}B_{x}-\widetilde{J}_{x}\right]|\left(t+dt/2,x+dx/2, y, z\right),\\
&\frac{dE_y}{dt}\left(t+dt/2,x, y+dy/2, z\right)=\left[\partial_{z}B_{x}-\partial_{x}B_{z}-\sigma\,E_{y}-\sigma_{\chi}B_{y}-\widetilde{J}_{y}\right]|\left(t+dt/2,x, y+dy/2, z\right),\\
&\frac{dE_z}{dt}\left(t+dt/2,x, y, z+dz/2\right)=\left[\partial_{x}B_{y}-\partial_{y}B_{x}-\sigma\,E_{z}-\sigma_{\chi}B_{z}-\widetilde{J}_{z}\right]|\left(t+dt/2,x, y, z+dz/2\right),\\
&\frac{dB_x}{dt}\left(t,x, y+dy/2, z+dz/2\right)=\left[\partial_{z}E_{y}-\partial_{y}E_{z}\right]|\left(t,x, y+dy/2, z+dz/2\right),\\
&\frac{dB_y}{dt}\left(t,x+dx/2, y, z+dz/2\right)=\left[\partial_{x}E_{z}-\partial_{z}E_{x}\right]|\left(t,x+dx/2, y, z+dz/2\right),\\
&\frac{dB_z}{dt}\left(t,x+dx/2, y+dy/2, z\right)=\left[\partial_{y}E_{x}-\partial_{x}E_{y}\right]|\left(t,x+dx/2, y+dy/2, z\right).
\end{split}
\end{align}
Let us use the center difference method at the given points to solve these equations. One can get the following results,
\begin{align}
\begin{split}
&E_{x}|^{n+1}_{i+1/2,j,k}=E_{x}|^{n}_{i+1/2,j,k}+dt \frac{B_{z}|^{n+1/2}_{i+1/2,j+1/2,k}-B_{z}|^{n+1/2}_{i+1/2,j-1/2,k}}{dy}-dt\frac{B_{y}|^{n+1/2}_{i+1/2,j,k+1/2}-B_{y}|^{n+1/2}_{i+1/2,j,k-1/2}}{dz}\\
&\qquad \qquad \qquad -dt\left\{\sigma\,E_{x}|^{n+1/2}_{i+1/2,j,k}+\sigma_{\chi}B_{x}|^{n+1/2}_{i+1/2,j,k}+\widetilde{J}_{x}|^{n+1/2}_{i+1/2,j,k}\right\},\\
&E_{y}|^{n+1}_{i,j+1/2,k}=E_{y}|^{n}_{i,j+1/2,k}+dt\frac{B_{x}|^{n+1/2}_{i,j+1/2,k+1/2}-B_{x}|^{n+1/2}_{i,j+1/2,k-1/2}}{dz}-dt\frac{B_{z}|^{n+1/2}_{i+1/2,j+1/2,k}-B_{z}|^{n+1/2}_{i-1/2,j+1/2,k}}{dx}\\
&\qquad \qquad \qquad -dt\left\{\sigma\,E_{y}|^{n+1/2}_{i,j+1/2,k}+\sigma_{\chi}B_{y}|^{n+1/2}_{i,j+1/2,k}+\widetilde{J}_{y}|^{n+1/2}_{i,j+1/2,k} \right\},\\
&E_{z}|^{n+1}_{i,j,k+1/2}=E_{z}|^{n}_{i,j,k+1/2}+dt\frac{B_{y}|^{n+1/2}_{i+1/2,j,k+1/2}-B_{y}|^{n+1/2}_{i-1/2,j,k+1/2}}{dx}-dt\frac{B_{x}|^{n+1/2}_{i,j+1/2,k+1/2}-B_{x}|^{n+1/2}_{i,j-1/2,k+1/2}}{dy}\\
&\qquad \qquad \qquad -dt\left\{\sigma\,E_{z}|^{n+1/2}_{i,j,k+1/2}+\sigma_{\chi}B_{z}|^{n+1/2}_{i,j,k+1/2}+\widetilde{J}_{z}|^{n+1/2}_{i,j,k+1/2} \right\}.
\end{split}
\end{align}
\begin{align}
\begin{split}
&B_{x}|^{n+1/2}_{i,j+1/2,k+1/2}=B_{x}|^{n-1/2}_{i,j+1/2,k+1/2}+dt\left\{ \frac{E_{y}|^{n}_{i,j+1/2,k+1}-E_{y}|^{n}_{i,j+1/2,k}}{dz}-\frac{E_{z}|^{n}_{i,j+1,k+1/2}-E_{z}|^{n}_{i,j,k+1/2}}{dy} \right\},\\
&B_{y}|^{n+1/2}_{i+1/2,j,k+1/2}=B_{y}|^{n-1/2}_{i+1/2,j,k+1/2}+dt\left\{ \frac{E_{z}|^{n}_{i+1,j,k+1/2}-E_{z}|^{n}_{i,j,k+1/2}}{dx}-\frac{E_{x}|^{n}_{i+1/2,j+1,k+1}-E_{x}|^{n}_{i+1/2,j,k}}{dz} \right\},\\
&B_{z}|^{n+1/2}_{i+1/2,j+1/2,k}=B_{z}|^{n-1/2}_{i+1/2,j+1/2,k}+dt\left\{ \frac{E_{x}|^{n}_{i+1/2,j+1,k}-E_{x}|^{n}_{i+1/2,j,k}}{dy}-\frac{E_{y}|^{n}_{i+1,j+1/2,k}-E_{y}|^{n}_{i,j+1/2,k}}{dx} \right\}.
\end{split}
\end{align}
These two sets of equations should be further simplified for the next numerical simulations. Then the electric parts can be cast into the following,
\begin{align}\label{eq:algorithm-E}
	\begin{split}
		&E_{x}|^{n+1}_{i+1/2,j,k}=COE|^{n+1/2}_{i+1/2,j,k}E_{x}|^{n}_{i+1/2,j,k}+CE|^{n+1/2}_{i+1/2,j,k}\Big[ \frac{dt}{dy}\left( B_{z}|^{n+1/2}_{i+1/2,j+1/2,k}-B_{z}|^{n+1/2}_{i+1/2,j-1/2,k} \right)\\
		&\qquad \qquad \qquad -\frac{dt}{dz}\left( B_{y}|^{n+1/2}_{i+1/2,j,k+1/2}-B_{y}|^{n+1/2}_{i+1/2,j,k-1/2} \right)-dt \sigma_{\chi}|^{n+1/2}_{i+1/2,j,k}B_{x}|^{n+1/2}_{i+1/2,j,k}-dt \widetilde{J}_{x}|^{n+1/2}_{i+1/2,j,k} \Big]\\
		&E_{y}|^{n+1}_{i,j+1/2,k}=COE|^{n+1/2}_{i,j+1/2,k}E_{y}|^{n}_{i,j+1/2,k}+CE|^{n+1/2}_{i,j+1/2,k}\Big[ \frac{dt}{dz}\left( B_{x}|^{n+1/2}_{i,j+1/2,k+1/2}-B_{x}|^{n+1/2}_{i,j+1/2,k-1/2} \right)\\
		&\qquad \qquad \qquad -\frac{dt}{dx}\left( B_{z}|^{n+1/2}_{i+1/2,j+1/2,k}-B_{z}|^{n+1/2}_{i-1/2,j+1/2,k} \right)-dt \sigma_{\chi}|^{n+1/2}_{i,j+1/2,k}B_{y}|^{n+1/2}_{i,j+1/2,k}-dt \widetilde{J}_{y}|^{n+1/2}_{i,j+1/2,k} \Big]\\
		&E_{z}|^{n+1}_{i,j,k+1/2}=COE|^{n+1/2}_{i,j,k+1/2}E_{z}|^{n}_{i,j,k+1/2}+CE|^{n+1/2}_{i,j,k+1/2}\Big[ \frac{dt}{dx}\left( B_{y}|^{n+1/2}_{i+1/2,j,k+1/2}-B_{y}|^{n+1/2}_{i-1/2,j,k+1/2} \right)\\
		&\qquad \qquad \qquad -\frac{dt}{dy}\left( B_{x}|^{n+1/2}_{i,j+1/2,k+1/2}-B_{x}|^{n+1/2}_{i,j-1/2,k+1/2} \right)-dt \sigma_{\chi}|^{n+1/2}_{i,j,k+1/2}B_{z}|^{n+1/2}_{i,j,k+1/2}-dt \widetilde{J}_{z}|^{n+1/2}_{i,j,k+1/2} \Big]\\
	\end{split}
\end{align}
Where we have used an approximation $E_{i}|^{n+1/2}=(E_{i}|^{n+1}+E_{i}|^{n})/2$ for the term $\sigma\,E_{i}$, herein $i=x,y,z$. 

While the magnetic field parts can be written as,
\begin{align}\label{eq:algorithm-B}
	\begin{split}
		&B_{x}|^{n+1/2}_{i,j+1/2,k+1/2}=B_{x}|^{n-1/2}_{i,j+1/2,k+1/2}+\frac{d t}{dz}\left( E_{y}|^{n}_{i,j+1/2, k+1}-E_{y}|^{n}_{i,j+1/2,k}  \right)\\
		&\qquad \qquad \qquad \qquad \qquad \qquad \qquad \qquad \qquad -\frac{d t}{dy}\left( E_{z}|^{n}_{i,j+1, k+1/2}-E_{z}|^{n}_{i,j,k+1/2}  \right)\\
		&B_{y}|^{n+1/2}_{i+1/2,j,k+1/2}=B_{y}|^{n-1/2}_{i+1/2,j,k+1/2}+\frac{d t}{dx}\left( E_{z}|^{n}_{i+1,j, k+1/2}-E_{z}|^{n}_{i,j,k+1/2}  \right)\\
		&\qquad \qquad \qquad \qquad \qquad \qquad \qquad \qquad \qquad -\frac{d t}{dz}\left( E_{x}|^{n}_{i+1/2,j, k+1}-E_{x}|^{n}_{i+1/2,j,k}  \right)\\
		&B_{z}|^{n+1/2}_{i+1/2,j+1/2,k}=B_{z}|^{n-1/2}_{i+1/2,j+1/2,k}+\frac{d t}{dy}\left( E_{x}|^{n}_{i+1/2,j+1, k}-E_{x}|^{n}_{i+1/2,j,k}  \right)\\
		&\qquad \qquad \qquad \qquad \qquad \qquad \qquad \qquad \qquad -\frac{d t}{dx}\left( E_{y}|^{n}_{i+1,j+1/2, k}-E_{y}|^{n}_{i,j+1/2,k}  \right)
	\end{split}
\end{align}
Firstly, the corresponding coordinates $(n,i,j,k)$ denotes $(t_0+n\,dt, x_0+i\,dx, y_0+j\,dy, z_0+k\,dz)$, herein $n,i,j,k=0,1,2,\cdots$. And the coefficients 
\begin{align}
	&CE|^{n}_{i,j,k}=\frac{2}{2+dt\,\sigma|^{n}_{i,j,k}},\qquad COE|^{n}_{i,j,k}=\frac{2-dt\,\sigma|^{n}_{i,j,k}}{2+dt\,\sigma|^{n}_{i,j,k}}.
\end{align}
In which $\sigma|^{n}_{i,j,k}=\sigma(t_0+n\,dt, x_0+i\,dx, y_0+j\,dy, z_0+k\,dz)$.

Secondly, the external source terms,
\begin{align}
	\begin{split}
		&\widetilde{J}_{m}|^{n}_{i,j,k}=\sigma|^{n}_{i,j,k}E^{ext}_{m}|^{n}_{i,j,k}+\sigma_{\chi}|^{n}_{i,j,k}B^{ext}_{m}|^{n}_{i,j,k},\qquad m=(x,y,z).
	\end{split}
\end{align}
Finally, the coordinates of the magnetic field at the last second term in each equation of the set of equations (\ref{eq:algorithm-E})) is not located at the same position of the computing magnetic field. So we need to represent it with the computing magnetic field. There is one method to solve this problem, i.e 
\begin{align}
	\begin{split}
		&B_{x}|^{n+1/2}_{i+1/2,j,k}=\frac{1}{8}\Big(B_{x}|^{n+1/2}_{i+1,j+1/2,k+1/2}+B_{x}|^{n+1/2}_{i+1,j+1/2,k-1/2}+B_{x}|^{n+1/2}_{i+1,j-1/2,k+1/2}+B_{x}|^{n+1/2}_{i+1,j-1/2,k-1/2} \\
		&\qquad \qquad \qquad \qquad \qquad B_{x}|^{n+1/2}_{i,j+1/2,k+1/2}+B_{x}|^{n+1/2}_{i,j+1/2,k-1/2}+B_{x}|^{n+1/2}_{i,j-1/2,k+1/2}+B_{x}|^{n+1/2}_{i,j-1/2,k-1/2}\Big),\\
		&B_{y}|^{n+1/2}_{i,j+1/2,k}=\frac{1}{8}\Big(B_{y}|^{n+1/2}_{i+1/2,j+1,k+1/2}+B_{y}|^{n+1/2}_{i+1/2,j+1,k-1/2}+B_{y}|^{n+1/2}_{i-1/2,j+1,k+1/2}+B_{y}|^{n+1/2}_{i-1/2,j+1,k-1/2} \\
		&\qquad \qquad \qquad \qquad \qquad B_{y}|^{n+1/2}_{i+1/2,j,k+1/2}+B_{y}|^{n+1/2}_{i+1/2,j,k-1/2}+B_{y}|^{n+1/2}_{i-1/2,j,k+1/2}+B_{y}|^{n+1/2}_{i-1/2,j,k-1/2}\Big),\\
		&B_{z}|^{n+1/2}_{i,j,k+1/2}=\frac{1}{8}\Big(B_{z}|^{n+1/2}_{i+1/2,j+1/2,k+1}+B_{z}|^{n+1/2}_{i+1/2,j-1/2,k+1}+B_{z}|^{n+1/2}_{i-1/2,j+1/2,k+1}+B_{z}|^{n+1/2}_{i-1/2,j-1/2,k+1} \\
		&\qquad \qquad \qquad \qquad \qquad B_{z}|^{n+1/2}_{i+1/2,j+1/2,k}+B_{z}|^{n+1/2}_{i+1/2,j-1/2,k}+B_{z}|^{n+1/2}_{i-1/2,j+1/2,k}+B_{z}|^{n+1/2}_{i-1/2,j-1/2,k}\Big),\\
	\end{split}
\end{align}

\section{Levi--Civita tensor and electromagnetic tensor in Milne space}\label{ap:levi-civita}\label{app-3}
In Milne coordinate, the Levi--Civita tensor is different from those in the Minkowski coordinate,
\begin{align}\begin{split}
		&\epsilon^{\mu\nu\rho\sigma}=\frac{1}{\sqrt{|g|}}\widetilde{\epsilon}^{\mu\nu\rho\sigma}
		,\qquad
		\epsilon_{\mu\nu\rho\sigma}=\theta(g)\sqrt{|g|}\widetilde{\epsilon}_{\mu\nu\rho\sigma}
		,
\end{split}\end{align}
where $g=\mathrm{det}(g_{\mu\nu})$, $\theta$ is the Heaviside step function, and the Levi--Civita symbol in the Minkowski coordinate $\widetilde{\epsilon}^{\mu\nu\rho\sigma}$ is defined by the following, 
\begin{align*}
	&\widetilde{\epsilon}^{\mu\nu\rho\sigma}=\widetilde{\epsilon}_{\mu\nu\rho\sigma}=\left\{
	\begin{array}{rl}
	    +1,&\qquad \text{even permutation of (0,1,2,...n-1)}, \\
		-1,&\qquad \text{odd permutation of (0,1,2,...n-1)},\\
	    0,&\qquad \text{otherwise}.
	\end{array}
	\right.
\end{align*}
Taking that $g=-\tau^{2}$ and $|g|=\tau^{2}$, one finds $\epsilon^{0123}=\frac{1}{\tau}$ and $\epsilon_{0123}=-\tau$.

The electric and magnetic fields in the Milne space are defined by the following,
\begin{align}
	&\widetilde{E}^{i}=F_{M}^{i0},\qquad \widetilde{B}^{i}=\widetilde{F}_{M}^{i0}.
\end{align}
Then one can directly derive the electromagnetic tensor in Milne space with the above Levi-Civita definition, which can be expressed as the following,
\begin{align}
	&F^{\mu\nu}_{M}=\begin{pmatrix}
		0 & -\widetilde{E}^{x} & -\widetilde{E}^{y} & -\widetilde{E}^{\eta}\\
		\widetilde{E}^{x} & 0 & -\tau\,\widetilde{B}^{\eta} &\frac{\widetilde{B}^{y}}{\tau}\\
		\widetilde{E}^{y} &\tau\,\widetilde{B}^{\eta} & 0 & -\frac{\widetilde{B}^{x}}{\tau} \\
		\widetilde{E}^{\eta} & -\frac{\widetilde{B}^{y}}{\tau} &\frac{\widetilde{B}^{x}}{\tau} &0 \\
	\end{pmatrix}, \qquad 	&\widetilde{F}^{\mu\nu}_{M}=\begin{pmatrix}
		0 & -\widetilde{B}^{x} & -\widetilde{B}^{y} & -\widetilde{B}^{\eta}\\
		\widetilde{B}^{x} & 0 & \tau\,\widetilde{E}^{\eta} &-\frac{\widetilde{E}^{y}}{\tau}\\
		\widetilde{B}^{y} &-\tau\,\widetilde{E}^{\eta} & 0 & \frac{\widetilde{E}^{x}}{\tau} \\
		\widetilde{B}^{\eta} & \frac{\widetilde{E}^{y}}{\tau} &-\frac{\widetilde{E}^{x}}{\tau} &0 \\
	\end{pmatrix}
\end{align}

\section{The velocity in Milne space and Minkowski space} \label{app:t-v}\label{app-4}
We will need some simplified velocities of the medium when we do the numerical calculations in the moving medium for doing comparisons, such as the static velocity, Bjorken velocity, and Gubser velocity. These velocities in the Milne space and Minkowski coordinate can be expressed as the following,
\begin{align}\begin{split}
		&u^{\mu}_{M}=R^{\mu}_{~\nu}u^{\nu}=\left\{\begin{aligned}
			&\left(\cosh\eta, 0, 0, -\frac{\sinh\eta}{\tau}\right),\qquad  \text{for static case: } u^{\mu}=(1,0,0,0),\\
			&\left(1, 0, 0, 0 \right),\qquad \qquad \qquad \qquad ~ \text{for Bjorken flow: } u^{\mu}=(\frac{t}{\tau},0,0,\frac{z}{\tau}),\\
			&\left(u^{\tau}, u^{\perp}\frac{x}{x_{\perp}}, u^{\perp}\frac{y}{x_{\perp}}, 0 \right),\qquad \text{for Gubser flow: } u^{\mu}=(u^{\tau}\,\cosh\eta,u^{\perp}\,\frac{x}{x_{\perp}}, u^{\perp}\,\frac{y}{x_{\perp}},u^{\tau}\,\sinh\eta).
		\end{aligned}
		\right.
\end{split}\end{align}
The velocity with M subscript represents the velocity in Milne coordinate, while the velocity without M subscript the velocity in Minkowski coordinate. The transformation matrix and anti-transformation matrix from Minkowski coordinate to Milne space are defined as follows,
\begin{align}
	&R^{\mu}_{~\nu}=\frac{\partial\,x^{\mu}_{M}}{\partial\,x^{\nu}}=\begin{bmatrix}
		\cosh\eta & 0 & 0 & -\sinh\eta\\
		0 & 1 & 0 & 0\\
		0 & 0 & 1 & 0 \\
		-\frac{\sinh\eta}{\tau} & 0 & 0 & \frac{\cosh\eta}{\tau}
	\end{bmatrix},\qquad \check{R}^{\mu}_{~\nu}=\frac{\partial\,x^{\mu}}{\partial\,x^{\nu}_{M}}=\begin{bmatrix}
		\cosh\eta & 0 & 0 & \tau\sinh\eta\\
		0 & 1 & 0 & 0\\
		0 & 0 & 1 & 0 \\
		\sinh\eta & 0 & 0 & \tau\cosh\eta
	\end{bmatrix}
\end{align}

The corresponding components in the velocity of the Gubser flow are expressed as the following~\cite{Gubser:2010ze, Gubser:2010ui},
\begin{align}
	&u^{\tau}=\frac{1+q^{2}\tau^{2}+q^{2}x_{\perp}^{2}}{2q\tau\sqrt{1+g^{2}}},\qquad u^{\perp}=\frac{qx_{\perp}}{\sqrt{1+g^{2}}},\qquad g=\frac{1+q^{2}x_{\perp}^{2}-q^{2}\tau^{2}}{2q\tau}.
\end{align}
Herein the proper time is defined by $\tau=\sqrt{t^{2}-z^2}$, and the transverse distance $x_{\perp}=\sqrt{x^{2}+y^{2}}$. While the temperature in the local rest frame of the fluid is defined as
\begin{align}
	&T=\frac{1}{\tau\,f^{1/4}_{*}}\left( \frac{\hat{T}_{0}}{(1+g^{2})^{1/3}}+\frac{H_{0}g}{\sqrt{1+g^{2}}}\left[1-(1+g^{2})^{1/6}{}_{2}F_{1}\left(\frac{1}{2}, \frac{1}{6}; \frac{3}{2}; -g^{2} \right) \right] \right).
\end{align}
$\hat{T}_{0}$ is a dimensionless integration constant, and $f_{*}=\epsilon/T^{4}=11$, $q=1/(4.3~\mathrm{fm})$. For RHIC energy $\sqrt{s_\mathrm{NN}}=200~\mathrm{GeV}$, $\hat{T}_{0}=5.55$ and $H_{0}=0.33$. The function ${}_{2}F_{1}$ denotes a hypergeometric function. 

For the Bjorken flow, the temperature changes as follows,
\begin{align}
	&T(\tau)=T_{0}\,\frac{\tau_{0}}{\tau},
\end{align}
where the $T_{0}$ is the temperature at time $\tau_{0}$, which can be given by the Glauber model. For example, $T_{0}\sim 400$ MeV at the center of the QGP at $\tau_{0}=0.4$ fm for $\sqrt{s_\mathrm{NN}}=200$ GeV Au+Au collisions. 
\end{widetext}

\end{appendix}

\bibliography{reference}
\bibliographystyle{unsrt}

\end{document}